\documentstyle[10pt]{article}
\def \susy {supersymmetry }
\def \brs {branching ratios }
\def \br {branching ratio }

\def\l {\lambda } 
 
\def \t {\theta }

\def\a {\alpha }

\def \d {\delta }

\def \g {\gamma }
\def \G {\Gamma }

\def \s {\sigma }
\def \e {\epsilon }
\def \ud {{1 \over 2} }

\def \bea {\begin{equation} }
\def \eea {\end{equation} }
\def \tchi  {{\tilde \chi} }
\def \Eslash {E \kern-.5em\slash }
\def \pslash {p \kern-.5em\slash }
\def \kslash {k \kern-.5em\slash }

\def \np {Nucl. Phys. }

\def \cc {coupling constant }
\def \ccs {coupling constants }

\title{Systematics of single superpartners
production at leptonic colliders
 \thanks 
{\it Supported by the 
 Laboratoire de la Direction des Sciences
de la Mati\`ere du Commissariat \`a l'Energie Atomique }    }
\author{{M. Chemtob and G. Moreau} \\ \\ \\
{\em  Service de Physique Th\'eorique }\\
{ \em  CE-Saclay F-91191 Gif-sur-Yvette,  Cedex FRANCE }}
\date{}

\if@twoside\oddsidemargin25mm
\else\oddsidemargin25mm\evensidemargin25mm\marginparwidth25mm\fi
\input psfig

\begin{document}
\maketitle

\begin{abstract}
We examine the effects of the lepton number violating R parity odd superpotential,
 $W= \ \l_{ijk} L_iL_jE_k^c$, on single production of fermion (charginos and neutralinos)
and scalar (sleptons and sneutrinos) superpartners at leptonic colliders for center of
 mass energies up to $500GeV \ -\  1TeV$. The probability 
amplitudes for all the five $2 \to 2$ body processes:
$l_J^+l_J^- \to \tchi_1^{\pm} l_m^{\mp}, \ \tchi_1^0 \nu_m  \ (\tchi_1^0 \bar \nu_m),
\  \tilde l_m^{\mp} W^{\pm}, \ \tilde \nu_m Z^0   (\tilde {\bar \nu}_m  Z^0) , 
\ \tilde \nu_m \g (\tilde {\bar \nu}_m  \g ) $, 
and the decays branching ratios for the produced superpartners are calculated at tree level.
The rates for all five reactions are proportionnal 
to $\lambda_{mJJ} ^2 $ where $J=1,2$ for 
$e^-e^+ $ and $\mu^-\mu^+$ colliders, respectively.
A semi-quantitative discussion is presented within a supergravity model assuming grand unification 
of gauge interactions and universal (flavor independent) soft supersymmetry 
breaking parameters $m_0$ (scalars), $m_{1/2}$ (gauginos) at the unification scale.
 The predictions obtained for 
the total and partial rates show that the single production reactions have a good potential
 of observability at the future $e^-e^+ $ and $\mu^-\mu^+$ supercolliders. 
For values of the
 R parity violating coupling constant of order $0.05$, the  $\tchi^{\pm,0}$ productions 
could probe all the relevant intervals for $\tan \beta $ and $m_0$ and
broad regions of the parameter space for the $\mu$ (Higgs mixing) and $m_{1/2}$ 
parameters ($\vert \mu \vert <400GeV, \ m_{1/2}<240GeV$),
while  the $\tilde \nu$ and $\tilde l$ productions could probe sneutrinos and sleptons masses up to
the kinematical limits ($m_{\tilde \nu}<500GeV, \ m_{\tilde l}<400GeV$). Using the hypothesis
 of a single dominant R parity violating coupling constant, a Monte Carlo events simulation
for the reactions, $l_J^+l_J^- \to \tchi_1^{\pm}l^{\mp}_m,\tchi_{1,2}^0 \nu_m,
\tchi_{1,2}^0 \bar \nu_m $,   
is employed to deduce some characteristic dynamical distributions of the final states.  
\end{abstract}


{T98/061}.......... {\it hep-ph/9807509} .... Phys. Rev. D: DU6463

\section{Introduction}

Should R parity turn out to be an approximate symmetry of
the minimal  supersymmetric
standard model, the truly quantitative tests of such a possibility would have to   
be sought in high energy colliders physics, as was first emphasized in
 \cite{Dimo,Barg,Ross}. The great majority of the existing theoretical
studies for the LEP or the Tevatron accelerators physics have focused on signals 
associated with the LSP (lightest supersymmetric particle)  decays and certain rare 
decays of the standard model particles (gauge \cite{Barg,Valle,Brahm,Lola} or Higgs \cite{Barg}
 bosons or top-quark \cite{Phill}). A few experimental searches have been attempted 
for $Z^0$ boson decays \cite{EX1,EX2}, for inos decays \cite{EX3,EX4}   and also in
more general settings \cite{EX5,EX6}.
Proceeding one step further, interesting proposals were made recently to explain the
so-called ALEPH anomalous four-jets events \cite{Aleph} on the basis of R parity violating 
decays of  neutralinos or charginos \cite{fourjet1,ghosh3,chank3}, squarks 
\cite{fourjet2,farrar1}, sleptons \cite{fourjet3}  or sneutrinos \cite{barger3} produced 
in pairs through the  two-body processes, $e^+ e^- \to
 \tchi^{0,+}  \tchi^{0,-} $ or $e^+e^- \to \tilde f \tilde {\bar f}$. 
(See \cite{expupdate} for recent updates and lists of references.)

Apart from precursor studies devoted to the HERA collider \cite{Butt,perez,aid},   little consideration was given  
in the past to single production of supersymmetric particles in spite
of the potential interest of a discovery of 
supersymmetry that might be accessible at 
 lower incident energies.
 The reason, of course, is the lack of information about the size of
 the R parity odd coupling constants other than the large number of 
indirects bounds deduced from low and intermediate energy phenomenology \cite{Bhatt}.
Therefore, for obvious reasons, the existing single production studies have rather
 focused on resonant production of sneutrinos, charged sleptons 
\cite{Dimo,Barg,EX5,Esmail,Zhang1,Zhang2,Feng,Kal1,Kal2}
or squarks \cite{Ross,Butt,Esmail,Zhang1,Zhang2}. The interpretation of the 
anomalous high $Q^2$ events recently observed at HERA by the ZEUS
 \cite{Zeus} and H1 \cite{H1} Collaborations, in terms of squark resonant production,
 has also stimulated a renewed interest in
 R parity violation phenomenology \cite{anomhera}.

The collider physics tests of supersymmetric models without R parity entail an important change
 in focus with respect to the conventional tests: degraded missing energy, diluted
 signals, additional background from the minimal supersymmetric standard model interactions and uncertainties 
from the R parity violation coupling constants compounded with those from the superpartners
 mass spectra. Our purpose in this work is to discuss semi 
quantitatively the potential for a discovery and the tests of supersymmetry with $2 \to 2$ body 
single superpartner production. Although several order of magnitudes in rates
 are lost with respect to the resonant single production, one can dispose here
  of a rich variety of phenomena with multilepton final states non diagonal in flavor.
 Besides, one may also test larger ranges of the sneutrino mass since this need not be 
restricted by the center of mass energy value.
Encouraged by the recent developments on R parity violation and by the prospects
of high precision measurements  at supercolliders \cite{Pesk},
 we propose to study single production at leptonic (electron and muon) 
colliders for the set of five $2 \to 2$ body reactions,
$l_J^+l_J^- \to \tchi^{\pm}l_m^{\mp}$,
$l_J^+l_J^- \to \tchi^0 \nu_m \ (\tchi^0 \bar \nu_m)$,
$l_J^+l_J^- \to \tilde  l_m^{\mp} W^{\pm} $,
$l_J^+l_J^- \to \tilde \nu_{mL} Z^0 \ (\tilde  {\bar \nu}_{mL} Z^0)$,
$l_J^+l_J^- \to \tilde \nu_{mL} \gamma 
\ (\tilde  {\bar \nu}_{mL} \gamma) \ [J=1,2],$
 in a more systematic way 
than has been attempted so far.
We limit ourselves to the lowest inos eigenstates.  
Let us note here that precursor indicative studies 
of the inos single production reactions were already presented in 
\cite{DESY1,DESY2} and that recent discussions concerning the reactions, 
$e^{\pm} \gamma  \to e^{\pm} \tilde \nu$ and $e^\pm \gamma 
\to \tilde l^{\pm} \nu$,
where the photon flux is radiated by one of the two beams, were presented in  
\cite{All}. We shall restrict our study to the lepton
 number violating interactions $L_iL_jE_k^c$ in association with 
the familiar gauge and Yukawa couplings of the minimal supersymmetric
standard model. The final states consist then of multileptons with or without 
hadronic jets. 
 
This paper contains four sections.
In section \ref{sectionF}, we present the main formalism for superpartners production cross
sections and decay rates. In section \ref{sectionR}, based on the supergravity approach
to supersymmetry soft breaking parameters, we present numerical results for the total
rates and the various branching ratios in wide regions of the parameter space. In section \ref{sectionCD},
 we show results for final states distributions of the processes,
 $l_J^+l_J^- \to \tchi^{\pm} l^{\mp},\tchi^0 \nu,\tchi^0 \bar \nu$, obtained by means 
of a Monte Carlo events simulation,
using the SUSYGEN routine \cite{Kats}. In section \ref{sectionC}, we state our conclusions.   

\section{General Formalism}

\label{sectionF}

Five $2 \to 2$ body single production reactions may be observed
at leptonic colliders. We shall use the following short hand notation to denote 
the associated probability amplitudes:
\begin{eqnarray}
M(\tchi_a^-+l_m^+)&=& M(l^-_J+l^+_J\to \tchi_a^- +l^+_m ), \cr 
M(\tchi_a^0+\bar \nu _m)&=& 
M(l^-_J+l^+_J \to \tchi_a^0 +\bar \nu_m ),  \cr 
M(\tilde l_{mL}^- +W^+)&=&
M(l^-_J+l^+_J\to  \tilde l_{mL}^- +W^+ ),\cr
M(\tilde \nu_m +Z)&=&
M(l^-_J+l^+_J\to  \tilde \nu_m + Z), \cr
M(\tilde \nu_m +\g )&=&
M(l^-_J+l^+_J\to  \tilde \nu_m +\g ),
\label{eqrc1}
\end{eqnarray}
where $J= 1,2$ is a flavor index for the initial state leptons (electrons 
and muons, respectively), 
the index  $a$  labels the charginos or neutralinos 
eigenvalues and  the index $ m $ the sleptons or sneutrinos families.
Our theoretical framework is the minimal supersymmetric standard model supplemented by the lepton number violating
R parity odd superpotential, $W= \ud \sum_{ijk} \l_{ijk} L_iL_jE_k^c$. This 
yields the sfermion-fermion Yukawa interactions,
\begin{eqnarray}
L&=&  \ud \sum_{[i\ne j, k]=1}^3 \l_{ijk}[\tilde  \nu_{iL}\bar e_{kR}e_{jL} +
\tilde e_{jL}\bar e_{kR}\nu_{iL} + \tilde e^\star _{kR}\bar \nu^c_{iR} e_{jL}
-(i\to j) ] +h.c.
\label{eqrc2}
\end{eqnarray}
where the sums labelled by indices, $i,j,k,$ run over the three 
leptons and neutrinos families with the condition $i\ne j$ following from the 
antisymmetry property, $\l_{ijk}=-\l_{jik}$..  

\subsection{Production Cross Sections}

\begin{figure}
\begin{center}
\leavevmode
\psfig{figure=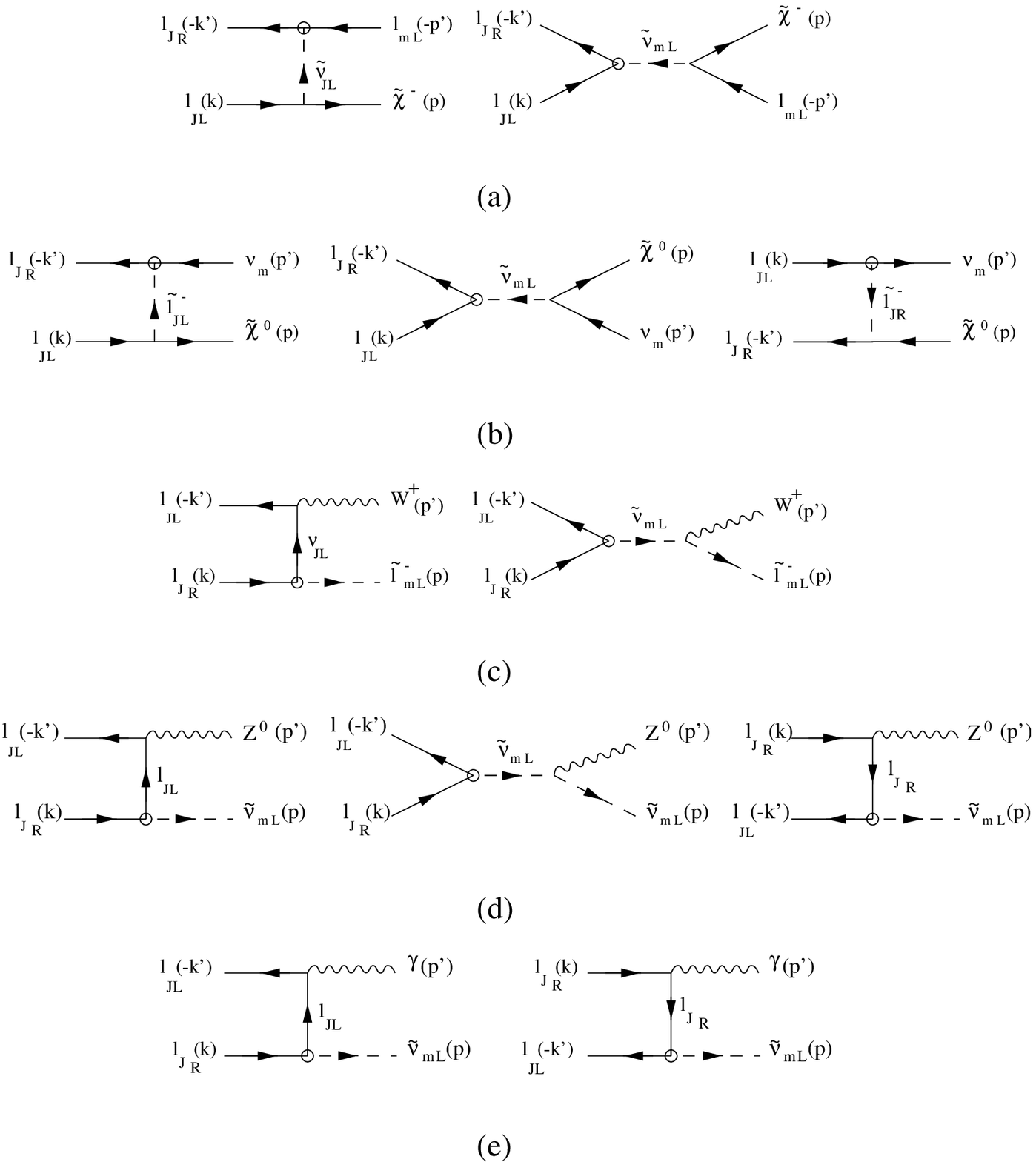}
\end{center}
\caption{\footnotesize  \it
Feynman diagrams for the processes, $l_J^+l_J^- \to \tchi^- l_m^+$ (a), 
 $l_J^+l_J^- \to \tchi^0 \bar \nu_m$ (b),  $l_J^+l_J^- \to \tilde l^-_{mL} W^+$ (c),
 $ l_J^+l_J^- \to \tilde \nu_{mL} Z^0 $ (d) and  $ l_J^+l_J^- 
 \to \tilde \nu_{mL} \ \g $ (e). 
The circled vertex correspond to the RPV interaction, with the coupling constant  $\l_{mJJ}$, and
the arrows denote flow of momentum.
\rm \normalsize }
\label{fig1}
\end{figure}

Each of the processes in eq.(\ref{eqrc1}) has a charge conjugate partner such that the
transformation between pairs of conjugate amplitudes can be formally described by
applying a CP transformation to the S-matrix. The relationship is most 
easily described at the level of
the amplitudes squared obtained after summation over the spins. Because of the simple 
action of CP on the initial state, $l_J^+(k)l_J^-(k')$, it can be seen that 
the amplitudes for the pairs of charge conjugated processes are related by the 
substitutions, $k \leftrightarrow k'$ 
and  $\l_{ijk} \leftrightarrow \l_{ijk} ^\star $.
The  tree level probability amplitudes   
are easily calculated by inspection of the Feynman diagrams given in
Fig.\ref{fig1}.    
The formulas for the amplitudes are consigned in
Appendix \ref{seca}.  
A few observations are in order at this point. First, the same configurations of 
lepton flavor indices, namely, $\l_{mJJ}$ with $J=1, \ m=2,3$  for $e^-e^+$ 
colliders and $J=2, \ m=1,3$  for $\mu^-\mu ^+$ colliders,
occur in all cases. Second, the amplitude for   
right chirality slepton $\tilde l_{mR}$ production has not been included in the 
above list of formulas for the reason that this is proportional to the  coupling constants
$\l_{11m}$ which vanishes by the antisymmetry property, $\l_{ijk}=-\l_{jik}$. 
Thirdly, all five processus can appear only in a single helicity configuration 
for the initial fermions (assumed massless), corresponding to identical helicities,
 namely, either $l^+_L l^-_L$ or $l^+_R l^-_R$ (recall that physical helicity 
for antiparticle is opposite to chirality).
Lastly, we observe that the relative signs between the $s,t$ and $u$ channels 
contributions are dictated by  both the structure of the interaction
Lagrangian and the signs of the Wick contractions for fermions. The results
for the spin summed squared amplitudes are given by somewhat
 complicated formulas which we have
 assembled in Appendix \ref{seca}. We have checked that our formulas for $\tchi^0$ and 
 $\tchi^{\pm}$ productions agree with the results provided
 in \cite{DESY1,DESY2} and \cite{Kats1}.

\subsection{Decays}

\label{sectionD}

\begin{table}[t]
\begin{center}
\begin{tabular}{|c|c|c|c|}
\hline
 Mass Intervals & Decays & Final State &  $\l_{m11}$\\ 
\hline 
& & & \\
$m_{\tilde l^-} > m_{\tchi^-}    $ (1) &  $\bullet  \tchi^-\to 
\bar \nu_i \bar \nu_j l_k$ & (A)$l_m^+ l_k^-\Eslash $ & $l^+_m e^-$ \\
$m_{\tilde l^-} < m_{\tchi^-}   $ (2)&   $\bullet  
\tchi^-\to \bar \nu_i \tilde l^-_i \to \bar \nu_i l_k \bar \nu_j  $ \ & & \\ 
\hline
& & & \\ 
$ m_{\tilde \nu }> m_{\tchi^-}   $ (3)&  $\bullet  
\tchi^-\to  l_j \bar l_k l_i $ &(B)$l_m^+ l_k^+  l_i^- l_j^- $ & $l^+_ml^-_me^+e^-$ \\   
$ m_{\tilde \nu }< m_{\tchi^-}   $  (4)&  $\bullet
\tchi^-\to l_i \tilde {\bar \nu}_i   \to  l_i l_j \bar l_k $& & \\ 
\hline
& & & \\ 
$m_{\tilde l},m_{\tilde \nu} > 
m_{\tchi^-} > m_{\tchi^0}   $  (5)&  $\bullet
\tchi^-\to \tchi^0 l_p \bar \nu_p  \to l_p \bar \nu_p \nu_i l_j\bar l_k $ & (C)$l_m^+ l^-_p  
l_k^{\pm}  l^{\mp}_i\Eslash $& $l^+_ml^-_pe^+e^-,$ \\
$m_{\tchi^-}  > m_{\tilde l}>m_{\tchi^0}  $  (6)& $\bullet \tchi^-\to \bar \nu_p
\tilde l_p^- \to   \bar \nu_p l_p \tchi^0  $& & $ l^+_ml^-_pe^{\pm}l_m^{\mp}$\\
 & $  \to  l_p \bar \nu_p \nu_i l_j \bar l_k $ & & \\
$m_{\tchi^-} > m_{\tilde \nu }>m_{\tchi^0}   $  (7) & $\bullet  \tchi^-\to
 l_p  \tilde{\bar \nu}_p   \to  l_p \bar \nu_p  \tchi^0
$ & & \\
& $  \to l_p \bar \nu_p \nu_i   l_j\bar l_k $ & & \\ 
\hline
& & & \\ 
$m_{\tilde q} > 
m_{\tchi^-} > m_{\tchi^0}   $ (8)&  $\bullet
\tchi^-\to \tchi^0 q_p \bar q_p, \  
 \to  q_p \bar q_p \nu_i l_j\bar l_k  $ & (D)$l_m^+ l_k^{\pm} l^{\mp}_i
 \Eslash +2jet$ & $l^+_me^+e^-,$ \\
$m_{\tchi^-}  > m_{\tilde q}>m_{\tchi^0}  $ (9) & $\bullet  \tchi^-\to 
\bar q_p \tilde q_p, \   \to q_p \bar q_p  \tchi^0  $& &$ l^+_me^{\pm}l_m^{\mp}$ \\
& $  \to  q_p \bar q_p \nu_i l_j \bar l_k $ & & \\ 
\hline 
& & & \\
$m_{\tchi^-} > m_{\tchi^0} +m_W  $ (10) & $\bullet
\tchi^-\to \tchi^0 W^-  \to  W^- \nu_i l_j\bar l_k $ & (E)$ l_m^+
l_k^{\pm} l^{\mp}_i W^-\Eslash $ & $l^+_me^+e^-,$ \\
&&&$ \ l^+_me^{\pm}l_m^{\mp}$\\ 
\hline 
\end{tabular}
\caption{\footnotesize  \it
The allowed chargino decays for different relative orderings of the superpartners 
masses. The column fields give the mass intervals, the decay schemes, the 
final states corresponding to the process, $l_J^+l_J^- \to \tchi^- l^+_m$,
with a single dominant \cc $\l_{ijk}$
 and the leptonic components of the final states  in the case of a single 
dominant coupling constant $\l_{m11} \ [m=2,3]$. The notation $\Eslash $ stands for missing energy associated with neutrinos. \rm \normalsize}
\label{tabloC}
\end{center}
\end{table}
 
\begin{table}[t]
\begin{center}
\begin{tabular}{|c|c|c|c|}
\hline
Mass Intervals & Decays & Final State &  $\l_{m11}$\\
\hline
& & & \\
$m_{\tilde \nu} < m_{\tchi^+}   $ (1) &  $ \bullet
\tilde \nu_m \to l_k \bar l_j $& (A)$l^-_k l^+_jZ^0$ & $e^+e^-$ \\
\hline
& & & \\
$m_{\tilde \nu} > m_{\tchi^+}   $ (2) &  $ \bullet
\tilde \nu_{m} \to l_{m} \tchi^+ \to l_m \bar l_i \bar l_j l_k$&
(B)$l^+_i l^+_j l^-_k l_{m}^- Z^0$ & $e^+e^-l^+_ml^-_m$ \\
\hline
& & & \\
$m_{\tilde \nu},\ m_{\tilde l} > m_{\tchi^+}   $ (3) &  $ \bullet
\tilde \nu_{m} \to l_{m} \tchi^+ \to l_m \nu_i \nu_j \bar l_k$&(C)
$l_k^+ l^-_{m} \Eslash  Z^0$ & $e^+l_m^-$ \\
$m_{\tilde \nu} > m_{\tchi^+} > m_{\tilde l} $ (4) &  $ \bullet
\tilde \nu_{m} \to l_{m} \tchi^+ \to l_m\tilde l^+_j \nu_j$
& & \\
& $ \to l_m \nu_j \nu_i \bar l_k$ & & \\
\hline 
& & & \\
$m_{\tilde \nu} > m_{\tchi^0}   $ (5)&  $ \bullet
\tilde \nu_m \to \nu_m \tchi^0 \to \nu_m \nu_i l_j \bar 
l_k $&(D)$ l_k^{\pm} l^{\mp}_i\Eslash Z^0$ & $e^+e^-,$ \\
&&& $ \ e^{\pm}l_m^{\mp}$ \\
\hline
& & & \\
$m_{\tilde \nu} > m_{\tchi^+} > m_{\tchi^0}   $  (6) & $ \bullet 
\tilde \nu_{m} \to \tchi^+ l_{m} \to l_m \tchi^0 \bar l_p \nu_p$ &
(E)$l_p^+ l_{m}^- l_k^{\pm} l^{\mp}_i\Eslash Z^0$ & $l^+_p l^-_m e^+e^-,$ \\
& $   \to l_m \bar l_p \nu_p \nu_i l_j \bar l_k$ & &  $  l^+_p l^-_m e^{\pm} l_m^{\mp}$ \\
$m_{\tilde \nu} > m_{\tchi^+}  > m_{\tilde l}>m_{\tchi^0}  $  (7)&  $ \bullet
\tilde \nu_{m} \to \tchi^+ l_{m} \to l_m \nu_p \tilde l_p^+$ & & \\
&  $  \to l_m \nu_p \bar l_p \tchi^0 \to l_m \nu_p \bar l_p \nu_i l_j \bar l_k $ & & \\ 
\hline
$m_{\tilde \nu} > m_{\tchi^+} > m_{\tchi^0}  $ (8) &$  \bullet  
\tilde \nu_{m} \to \tchi^+ l_{m} \to l_m \tchi^0 q_p \bar q_p$ &
(F) $ l^-_{m} l^{\pm}_k l^{\mp}_i Z^0+ 2jet $ & $l^-_me^+e^-,$ \\
& $  \to l_m q_p \bar q_p \nu_i l_j \bar l_k $ & & $  l^-_me^{\pm}l_m^{\mp}$\\
$m_{\tilde \nu} > m_{\tchi^+}  > m_{\tilde q}>m_{\tchi^0}  $ (9) & $ \bullet  
\tilde \nu_{m} \to \tchi^+ l_{m} \to l_m \bar q_p \tilde q_p  $ & & \\
& $ \to l_m \bar q_p q_p \tchi^0 \to l_m q_p \bar q_p \nu_i l_j \bar l_k $ & & \\
\hline
$m_{\tilde \nu} > m_{\tchi^+} > m_{\tchi^0}+m_W $ (10) & $ \bullet   
\tilde \nu_{m} \to \tchi^+ l_{m} \to l_m \tchi^0 W^+$ & (G)
$l^-_{m} l^{\pm}_k l^{\mp}_i \Eslash   W^+ Z^0$ & $l^-_me^+e^-,$ \\
& $  \to l_m W^+ \nu_i l_j \bar l_k $ & & $ l^-_me^{\pm}l_m^{\mp}$\\
\hline
\end{tabular}
\caption{\footnotesize  \it
The allowed sneutrino decays for different relative orderings of the superpartners 
masses. The column fields give the mass intervals, the decay schemes, the 
final states corresponding to the process, $l_J^+l_J^- \to Z^0 \tilde \nu_m$,
with a single dominant \cc $\l_{ijk}$
 and the leptonic components of the final states  in the case of a single 
dominant coupling constant $\l_{m11} \ [m=2,3]$. The notation $\Eslash $ stands for missing energy associated with neutrinos.
\rm \normalsize}
\label{tabloA}
\end{center}
\end{table}

\begin{table}[t]
\begin{center}
\begin{tabular}{|c|c|c|c|}
\hline
Mass Intervals & Decays & Final State &  $\l_{m11}$ \\
\hline
$m_{\tilde l^-} < m_{\tchi^-}   $ (1) &  $ \bullet
\tilde l^-_m\to l_k \bar \nu_i $& (A)$l^-_k \Eslash W^+$ & $e^-$ \\
\hline
$m_{\tilde l^-} > m_{\tchi^-}   $ (2) &  $ \bullet
\tilde l_m^-\to \nu_m \tchi^- \to \nu_m l_k \bar \nu_j \bar \nu_i$&
(B)$l^-_k \Eslash W^+$ & $e^-$ \\
\hline
$m_{\tilde l^-},\ m_{\tilde \nu} > m_{\tchi^-}   $ (3) &  $ \bullet
\tilde l_m^-\to \nu_m \tchi^- \to \nu_m l_j \bar l_k l_i$&(C)$l_k^+ l^-_i
 l_j^- \Eslash W^+$ & $e^+e^-l_m^-$ \\
$m_{\tilde l^-} > m_{\tchi^-} > m_{\tilde \nu}  $ (4) &  $ \bullet
\tilde l_m^-\to \nu_m \tchi^- \to \nu_m \tilde {\bar \nu}_i l_i $& & \\
& $ \to \nu_m l_i l_j \bar l_k $ &&\\
$m_{\tilde l^-} > m_{\tchi^0}   $ (5)&  $ \bullet
\tilde l_m^-\to l_{m} \tchi^0 \to l_m \nu_i l_j \bar 
l_k $&(D)$l_{m}^- l_k^{\pm} l^{\mp}_i\Eslash W^+$ & $l^-_me^+e^-,$ \\
&&& $ \ l^-_me^{\pm}l_m^{\mp}$ \\
\hline
$m_{\tilde l^-} > m_{\tchi^-} > m_{\tchi^0}   $  (6) & $ \bullet
\tilde l_m^- \to \tchi^- \nu_m \to \nu_m \tchi^0 l_p \bar \nu_p$ &
(E)$l_p^- l_k^{\pm} l^{\mp}_i\Eslash W^+$ & $l^-_pe^+e^-,$ \\
& $ \to \nu_m l_p \bar \nu_p \nu_i l_j \bar l_k $ &&$ l^-_pe^{\pm}l_m^{\mp}$  \\
$m_{\tilde l^-} > m_{\tchi^-}  > m_{\tilde \nu}>m_{\tchi^0} $  (7)&  $ \bullet 
\tilde l_m^- \to \tchi^- \nu_m \to \nu_m l_p \tilde {\bar \nu}_p$ 
& & \\
& $ \to \nu_m l_p \bar \nu_p \tchi^0 \to \nu_m l_p \bar \nu_p \nu_i l_j \bar l_k$ 
& & \\ 
\hline
$m_{\tilde l^-} > m_{\tchi^-} > m_{\tchi^0}  $ (8) & $\bullet  
\tilde l_m^- \to \tchi^- \nu_m \to \nu_m \tchi^0 q_p \bar q_p$ & (F)
$ l^{\pm}_k l^{\mp}_i \Eslash W^++ 2jet$ & $e^+e^-,$ \\
 & $ \to \nu_m q_p \bar q_p \nu_i l_j \bar l_k $ & & $ e^{\pm}l_m^{\mp}$  \\
$m_{\tilde l^-} > m_{\tchi^-}  > m_{\tilde q}>m_{\tchi^0}  $ (9) & $\bullet 
\tilde l_m^- \to \tchi^- \nu_m \to \nu_m \bar q_p \tilde q_p $ & & \\
& $\to \nu_m \bar q_p q_p \tchi^0 \to \nu_m \bar q_p q_p  \nu_i l_j \bar l_k $ & & \\
\hline
$m_{\tilde l^-} > m_{\tchi^-} > m_{\tchi^0}+m_W  $ (10) & $\bullet  
\tilde l_m^- \to \tchi^- \nu_m \to \nu_m \tchi^0 W^-$ & (G)
$l^{\pm}_k l^{\mp}_i W^-  \Eslash   W^+$ & $e^+e^-,$ \\
& $ \to \nu_m W^- \nu_i l_j \bar l_k $ & & $ e^{\pm}l_m^{\mp}$ \\
\hline 
\end{tabular}
\caption{\footnotesize  \it
The allowed slepton decays for different relative orderings of the superpartners 
masses. The column fields give the mass intervals, the decay schemes, the 
final states corresponding to the process, $l_J^+l_J^- \to W^+ \tilde l_m^-$,
with a single dominant \cc $\l_{ijk}$
 and the leptonic components of the final states  in the case of a single 
dominant coupling constant $\l_{m11} \ [m=2,3]$. The notation $\Eslash $ stands for missing energy associated with neutrinos.
\rm \normalsize}
\label{tabloB}
\end{center}
\end{table}

In order to exhibit the possible physical final states, we need to consider
 the decays of 
the produced  supersymmetric particles, taking into account both the  
minimal supersymmetric standard model interactions (denoted RPC or R parity  conserving) and the 
R parity odd interactions (denoted RPV or R parity  violating). \\
A number of hypotheses and approximations, which we list below, will be
 employed in the evaluation of partial rates.

\begin{itemize}
\item[1)] Supersymmetric particles  decays are assumed to have
narrow widths  (compared to their masses) and 
are produced on-shell with negligible spin correlations 
between the production and decay stages.
This allows us to apply the familiar  phase space
factorisation formula for the production cross sections.
\item[2)] Spin correlations are neglected at all stages of the cascade decays
 such that the branching ratios in single or double cascades can be obtained by applying recursively 
the standard factorisation formula.
\item[3)] Sleptons belonging to all  three families  and squarks  
belonging to the first two families  are degenerate in mass.
 Therefore, for a given decay process as, for instance, $\tchi^- \to \tilde
 {\bar \nu}_p l_p$, either all three generations will be energetically
 allowed or forbidden. Furthermore, flavor off-diagonal channels 
such as, $\tilde l_1 \to \tilde l_2+Z^0$,... are closed.
 \item[4)] The lowest eigenstates of neutralinos $\tchi_a^0$ and charginos
 $\tchi_a^{\pm}$ ($a=1$) are excited in the cascade chains. 
 \item[5)] All superpartners decay
 inside the detector volume. 
In the presence of broken R parity, the condition  
for  electric charge neutral LSPs  to decay inside the detector
yields comfortable  lower bounds  of order 
$\l > 10^{-7}$ \cite{Ross,Daw}.
\item[6)]  Either a single RPV coupling constant is dominant in both the production 
 and decay stages, or a pair of RPV coupling constants are dominant, one in the production
 stage ($\l_{mJJ}$) and the other in the decay stage ($\l_{ijk}$).
The latter case with two  dominant RPV \ccs may be of interest since strong bounds 
on quadratic products exist only for a few family configurations. The strongest
 bounds arise from the $\mu \to 3e$ decay \cite{quadbound}:
 $\l_{p11} \l_{p12}<6.5 \ 10^{-7},
 \  \l_{p21} \l_{p11}<6.5 \ 10^{-7} \ [p=2,3]$, while other quadratic product 
bounds are of 
order $10^{-3},10^{-4}$. Besides, as long as the coupling constant $\l$, which
controls the RPV decays, is small in comparison with the gauge coupling constants but not very much smaller (so that the LSP decays inside the detector), then the \br will
depend weakly on $\l$ since the last stage of the decay chain (LSP decay) is
 independent of $\l$.
\item[7)] The widths for the decays with four and higher body final states are neglected,
such as those which occur in slepton (sneutrino) decays for, $m_{\tchi^-},m_{\tchi^0}>
m_{\tilde l}$ ($m_{\tchi^-},m_{\tchi^0}>m_{\tilde \nu}$), mediated by virtual charginos
or neutralinos, namely, $\tilde l^-_m \to \nu_m l_k \bar \nu_j \bar \nu_i$ and
$\tilde {l^-}_m \to l_m \bar {l_k} l_j \nu_i $ ($\tilde {\nu}_m \to l_m l_k \bar {l_j} 
\bar {l_i} $ and $ \tilde {\nu}_m \to \nu_m \bar {l_k} l_j \nu_i)$.
\item[8)] A supergravity model for the soft \susy breaking parameters is used where, 
generically, $\tchi^0_1$ is the LSP.
\end{itemize}

The consideration of the various order relations in the superpartners mass spectrum
 leads to a list of decay schemes for the initially produced superparticles. These are
displayed in Tables \ref{tabloC},  \ref{tabloA} and \ref{tabloB}, for $\tchi^{\pm}$,
$\tilde \nu_L$ and $\tilde l_L$, respectively. The signals for the $\tchi^0_1$ decays
are very few in number and will be discussed separately in 
section \ref{neutralinosection}. 
Some comments on these tables are in order.
Except for hadronic dijet pairs from the decay processes, 
$\tchi^{\pm} \to \tchi^0 \bar q q'$, all other final particles will consists of
 multileptons and missing energy associated with neutrinos.
 In the hypothesis of a single dominant RPV coupling constant in the decays stage, 
one can deduce the various final states flavor configurations by an inspection
of the tables (cf. tables captions).
The produced $\tchi^0, \ \tchi^{\pm}, \ \tilde \nu, \ \tilde l^{\pm}$ will
 decay according to cascade schemes
 dictated by the superpartners mass spectrum. It is important to distinguish the 
direct RPV induced decays: $ \tchi^-\to \bar \nu_i \bar \nu_j l_k$,
 $ \tchi^-\to l_i  l_j \bar l_k$, 
$\tchi^0 \to \nu_i l_j  \bar l_k$, $\tchi^0 \to \bar \nu_i \bar l_j  l_k$,
 $\tilde \nu_i \to l_k \bar l_j$, $\tilde l_{kR}^-\to l_j \nu_i $,
 and  $\tilde l_{jL}^-\to l_k \bar \nu_i $, from
the indirect RPC induced decays: $\tilde l_L^- \to \tchi^- \bar \nu$, $\tilde \nu_L
 \to \tchi^+l$, $\tilde l^-_{L,R} \to \tchi^0  l$, $\tilde \nu_L \to \tchi^0 \nu$,
$\tchi^0 \to \tilde l^- {\bar l}$, $\tchi^0 \to l \tilde {l^+}$,
$\tchi^- \to \tilde l^- \bar \nu$, $\tchi^- \to l \tilde {\bar \nu}$ and $\tchi^{\pm} \to 
\tchi^0 W^{\pm}$ for two-body final states and, $\tchi^- \to \tchi^0 f \bar f$
 (f=leptons or quarks), for three-body final states.
All the formulas needed to evaluate the partial decay widths are quoted in the Appendix \ref{secappa}.
As can be seen from Tables \ref{tabloC},\ref{tabloA},\ref{tabloB}, 
a given final state can arise from different
processes, depending on the relative orderings of the masses. A reaction chain 
occuring through an intermediate particle which is produced on-shell leads obviously
 to the same final state when the production of this particle is
 kinematically forbidden and it must then occur through a virtual intermediate state.
In the approximation of family  degenerate sleptons, sneutrinos and squarks, 
the index $p$ in the tables runs over the three generations. A single exception
 is the hadronic decay, $\tchi^{\pm} \to \tchi^0 u_p \bar d_p (\tchi^0 d_p \bar u_p)$,
 which is restricted to the first two families because of the large top-quark mass.

Another subtle point concerns the multiplicity of a given signal, namely, the number of 
different configurations which can lead to the same final state. Due to the 
antisymmetry property of $\l_{ijk}$, the final states from chargino RPV 
decays (cf. A and B in Table \ref{tabloC}) have a multiplicity of two. The reason is that 
these decays proceed through the exchanges of the sleptons (or sneutrinos) in
 families $i$ and $j$, for a given $\l_{ijk}$.
This fact is already accounted for in the virtual $\tchi^{\pm}$ three-body decays 
(A(1),B(1),Table \ref{tabloC}), but must be put by hand in the $\tchi^{\pm}$ cascade decays 
proceeding through the on-shell production of sleptons or sneutrinos which decay
 subsequentially (A(2),B(2),Table \ref{tabloC}).

To get a better understanding of the interplay between RPC and RPV decays,
it is helpful to note that the \brs  can be written formally as,
 $B_D={\l^2 \over c \ g^2 +\l^2}$, for direct decays 
and, $B_I={g^2 B\over c_0 \ g^2 +\l^2}$, for two-stages indirect decays, where 
 $B={\l^2 \over c \ g^2 +\l^2}$ is the LSP branching ratio, $\l$ and $g$ are symbolic notations for the RPV 
and RPC coupling constants and $c,c_0$ are calculable constants.
Of course, $B=1$, if the last decaying particle is the LSP, 
which is the generical case. For values of the 
RPV coupling constants small with respect to the gauge coupling constants 
($\l \leq 0.05$), namely, $g^2 \gg \l^2$,
the dependence on $\l$ of the indirect decays \brs is weak and we have, $B_I \gg B_D$. For large enough RPV \cc (for example, $\l=0.1$) or for suppressed
indirect decays (due for example to kinematical reasons), the direct decays may become
competitive and both the direct and indirect \brs depend strongly on $\l$.
 
\section{The model and its parameter space}
\label{sectionmodel}

\begin{figure}
\begin{center}
\leavevmode
\psfig{figure=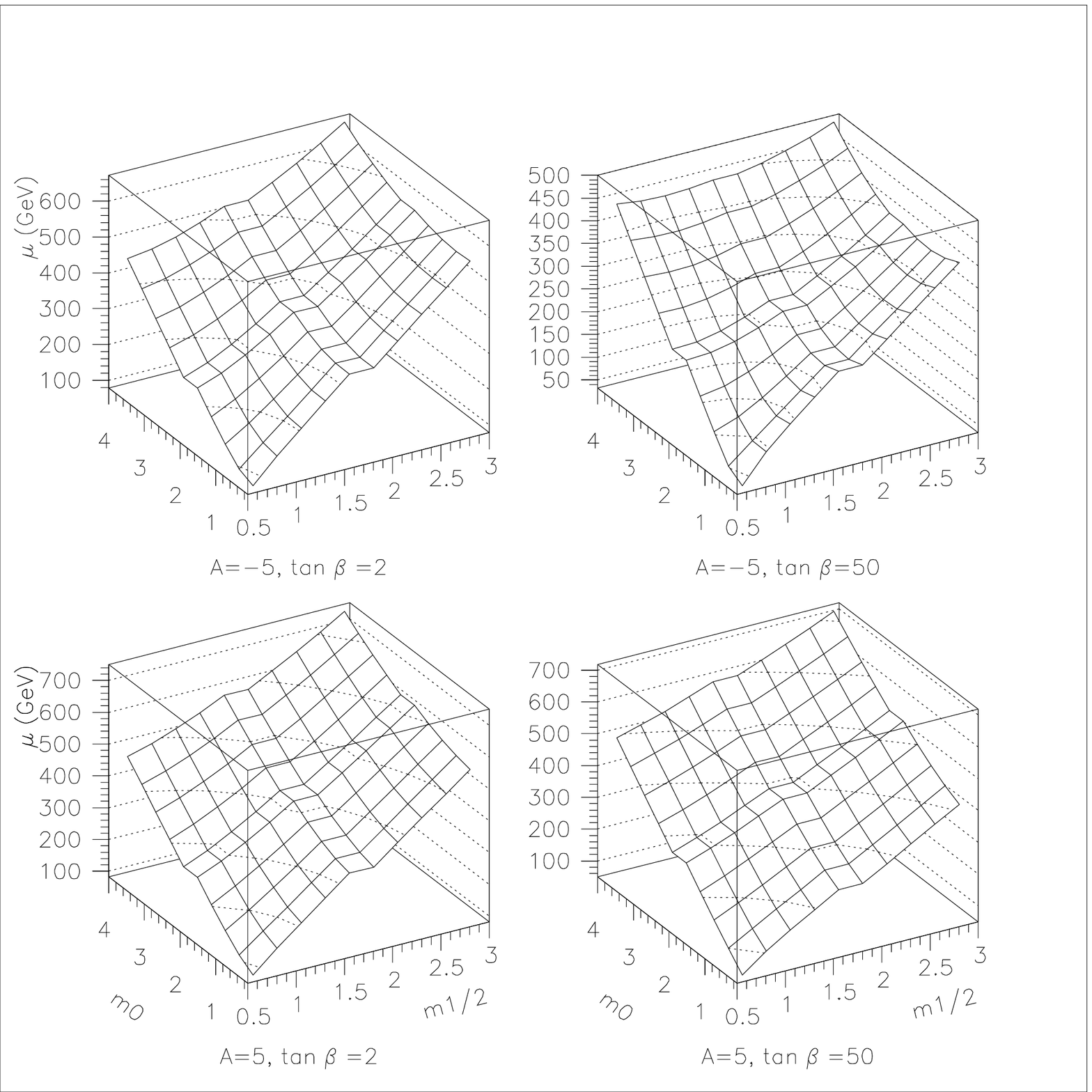}
\end{center}
\caption{\footnotesize  \it
The solution $\mu(t_Z)$, at scale $m_Z$, for the electroweak symmetry radiative breaking 
equations, at running top-quark mass, $m_{t}(m_{t})= 171  $ GeV 
$(m_t^{pole}= 180 $ GeV), is plotted as a function 
of ${m_0 \over 100GeV}$ and ${m_{1/2} \over 100GeV}$ for four 
values of the pair of  parameters, $A$ and $\tan \beta$.  
\rm \normalsize }
\label{efbr}
\end{figure}

\begin{figure}
\begin{center}
\leavevmode
\psfig{figure=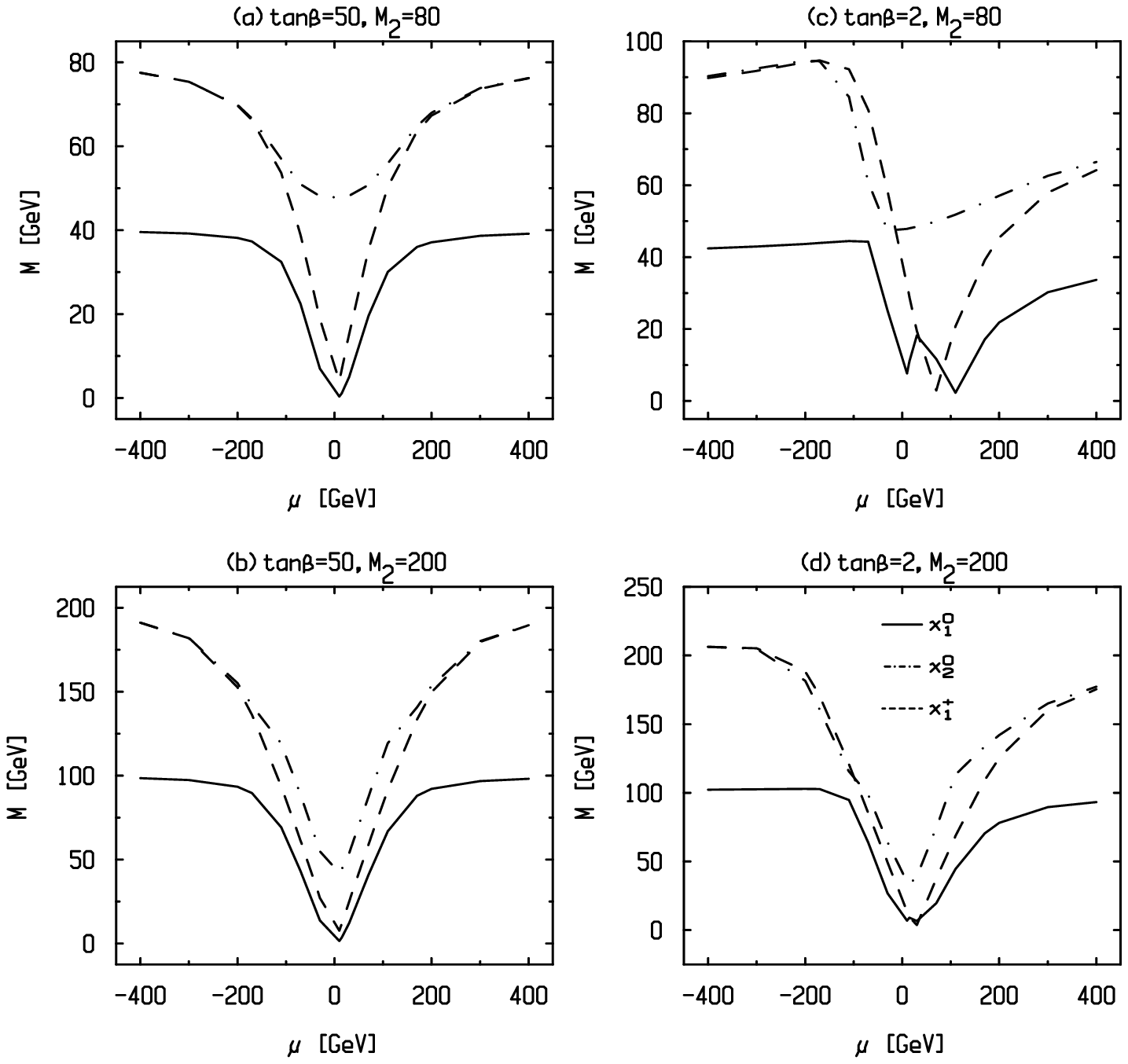}
\end{center}
\caption{\footnotesize  \it
Mass spectrum for the chargino $\tchi^{\pm}_1$ and the first 
two lowest mass neutralinos, $\tchi^0_1$ and $\tchi^0_2$, as a function of $\mu$.
Four choices of the parameters, $\tan \beta$ and $M_2$ (in $GeV$), are used, 
as indicated on top of each window. 
\rm \normalsize }
\label{spec}
\end{figure}

We shall develop the study of single superpartner production within a non minimal supergravity
framework, assuming the existence of a grand unified gauge theory and of family universal 
boundary conditions on 
the supersymmetry breaking parameters. The renormalization group improved classical spectrum
 of the scalar superpartners is determined in principle by the full set of soft supersymmetry breaking
 parameters at the unification scale, $M_X$, namely, $m_0$ (common scalars mass), $m_{1/2}$ 
(common gauginos mass), $A$ (trilinear Yukawa interactions), $B_{\mu}$ (bilinear Higgs
interaction); by the parameters $\tan \beta={v_u \over v_d }
={<H_u> \over <H_d>}$ and $ \mu(t)$, where $t$ denotes the running scale; and by the gauge 
coupling constants, $g_a(t)$, along with fermions masses, $m_f^2(t)$. 
If one neglects the Yukawa interactions of quarks and leptons with the Higgs bosons, then
the running masses of all sfermions remain family degenerate down to the electroweak breaking 
scale where they are described by the familiar additive formula,

\begin{eqnarray}
m_{\tilde f}^2(t)=m_{f}^2(t)+m_0^2+c_f(t)m_{1/2}^2 \pm m_{Z^0}^2 \cos (2 \beta) 
(T^f_3-Q^f x_W),
\label{specsugra}
\end{eqnarray}
  
where $c_f(t)$ are calculable coefficients depending on the gauge interactions parameters and the last term represents the D-term contribution,
the upper and lower sign being for the left and right chirality sfermions, respectively.
 The most relevant Yukawa coupling constants, namely, those of the third family of up-quarks or, for large $\tan \beta$, of d-quarks and leptons,
are expected to induce downwards shifts for the third family squarks (up and down) and sleptons,
which depend non trivially on the parameters $A$ and $\mu$. In this work, we shall restrict
consideration to the simple case of family independent running masses
and employ the 
approximate representation in eq.(\ref{specsugra}) with the numerical values 
quoted in \cite{Drees}.
Note that the total rates do not depend on the squarks masses and, as already remarked
in section \ref{sectionD}, the third families of squarks are not considered in the cascade decays.
The charginos and neutralinos classical mass spectra are determined by the subset
 of parameters: $M_1(t), \ M_2(t), \ \mu(t)$ and $\tan \beta$. For fixed $m_{1/2}$, 
the solution of the one loop renormalization group equations is given explicitly by, 
$m_{1/2} =  (1- \beta_a t) M_a(t),$
where  $t=\log({M_X^2 \over Q^2})$, $Q$ denoting the running scale,
$\beta_a = {g_X^2 b_a \over (4 \pi)^2 }$,
$b_a=(3,-1,-11)$ with $a=(3,2,1)$, corresponding to the
beta functions parameters for the  gauge group factors, 
$SU(3),SU(2)_L,U(1)_Y$, and $g_X$ is the coupling constant at unification
scale.
Note that the wino and bino masses are related as, $ M_1(t) = {5 \over 3}M_2(t) 
\tan^2 \theta _W $.
It is useful here to comment on the relation of our framework with the so-called minimal
supergravity framework in which one assumes a constrained parameter space compatible with 
electroweak symmetry breaking. Let us follow
here the so-called ambidextrous minimal supergravity
approach \cite{ewsb2}, where one chooses
$[m_0,m_{1/2},A, sign(\mu),\tan \beta]$ as the free parameters set and
 derives $\mu(t_Z),B_{\mu}(t_Z)$, at the electroweak symmetry breaking scale,
 $t_Z= \ln M_X^2/m_Z^2$, 
through the minimisation equations for the Higgs bosons potential. 
For fixed $m_0,m_{1/2}$ 
and $\tan \beta$, varying $A$ will let the parameter $\mu(t_Z)$ span finite intervals of
relatively restricted sizes. In figure \ref{efbr}, we  give results of 
a numerical resolution of the renormalization group equations which show
 the variation of $\vert \mu(t_Z) \vert $ as a function of $m_0$ and $m_{1/2}$, 
and also exhibit its dependence on $A$. Note that the equations admit the 
symmetry, $\mu (t_Z) \to - \mu  (t_Z)$.
Observing that  $\mu(t_Z)$ is typically  a monotonous increasing function of $A$, we see 
from Fig.\ref{efbr} that the  corresponding  incremental increase, 
$\d \mu (t_Z)/\mu (t_Z) $, as  one spans the wide interval, $A\in [-5, +5]$, 
is  small  and of order 20\%. 

In the  infrared fixed point  approach for the top-quark Yukawa coupling, 
$\tan \beta$ is fixed (up to the ambiguity associated with  large 
or low $\tan \beta$ solutions) in terms of the top-quark mass,
 $m_{t}=C \ \sin \beta, $ 
with, $C \simeq 190-210GeV$, for, $\alpha_3(m_{Z^0})=0.11-0.13$ \cite{Pok}.
The dependence on $A$ of
 the electroweak constraint also becomes very weak, so that 
  $\mu (t_Z)$ is a known function of $m_0,m_{1/2}$ and $\tan \beta$ \cite{Pok}:

\begin{eqnarray}
\mu^2+{m_{Z}^2 \over 2}=m_0^2{1+0.5\tan^2 \beta \over \tan^2\beta -1}+
m_{1/2}^2{0.5+3.5\tan^2\beta \over \tan^2\beta -1}.
\label{ewsb}
\end{eqnarray}

In section \ref{sectionBR}, we will discuss 
results for the branching ratios in this constrained model. The total rates are not affected in any
significant way by which version of the supergravity models is used, 
since, as we will see, their dependence on $\tan \beta$ and $m_0$ turns out to be smooth.

The main uncertain inputs are the superpartners mass spectrum and the coupling constants
$\l_{ijk}$. To survey the characteristic properties of single production over a broad
region of parameter space, we found it convenient to consider
a continuous interval of variation for $\mu(t_Z)$, namely, $\mu(t_Z) \in [-400,+400]GeV$, while 
choosing suitable discrete values for the other parameters: $M_2(t_Z)=50,80,100,150,200 \ GeV$,
$m_0=20,50,150 \ GeV$ and $\tan \beta=2,50$. We shall set the unification scale at 
$M_X=2 \ 10^{16}GeV$ and the running scale at $Q^2=m_Z^2$.
 For definiteness, we choose 
the coupling constant, which controls the size of the 
production cross section, at the reference value: $\l_{mJJ}=0.05$. This is the strongest
bound for a slepton mass of $100GeV$ \cite{Bhatt}. The dependence of integrated total
rates on $\l_{mJJ}$ is then given by a simple rescaling $({\l_{mJJ} \over 0.05})^2$ but
that of branching ratios on $\l_{ijk}$ (which may or may not be identified with $\l_{mJJ}$)
is more complicated because of the interplay between the RPC and RPV contributions which
add up in the total decay widths.
The reference value used here, $\l_{ijk}=0.05$, is also an interesting borderline
value since below this value the dependence of branching fractions on $\l_{ijk}$
becomes negligible in generic cases.

It will prove helpful in the following discussion to keep within 
sight the spectrum for the 
low-lying inos. We display in Figure \ref{spec} the results obtained by solving 
numerically the eigenvalues problem for the charginos and neutralinos mass matrices. 
Recall the current experimental bounds \cite{PartData}, $m_{\tchi^0_1}>23GeV$,
 $m_{\tchi^{\pm}_1}>45GeV$, $m_{\tilde \nu}>37.1GeV$ and $m_{\tilde l}>45GeV$. 
The following remarks about Figure \ref{spec} will 
prove relevant for the discussion on branching fractions: (i) The symmetry of the spectra
 under, $\mu \leftrightarrow - \mu$, is spoilt  at low $\tan \beta$ as can be seen on the 
explicit expression for the inos masses in \cite{haber}; (ii) The mass 
differences $\tchi^+-\tchi^0$ increase with $\vert \mu \vert$ with a steep rise appearing 
at, $\mu=M_2$, the borderline between the Higgsino and gaugino dominant regimes; (iii)
 The spacings $\tchi^0_2-\tchi^0_1$ and $\tchi^+_1-\tchi^0_1$ decrease in magnitudes,
 relatively to the $\tchi^0_1$ mass, with increasing $M_2$. Although we show here the 
results for $\tchi^0_2$ mass, the interesting possibility of exciting 
the second neutralino $\tchi^0_2$ is not considered in the subsequent discussion.

\section{Results and discussion}
\label{sectionR} 

\subsection{Total production rates}

\begin{figure}
\begin{center}
\leavevmode
\psfig{figure=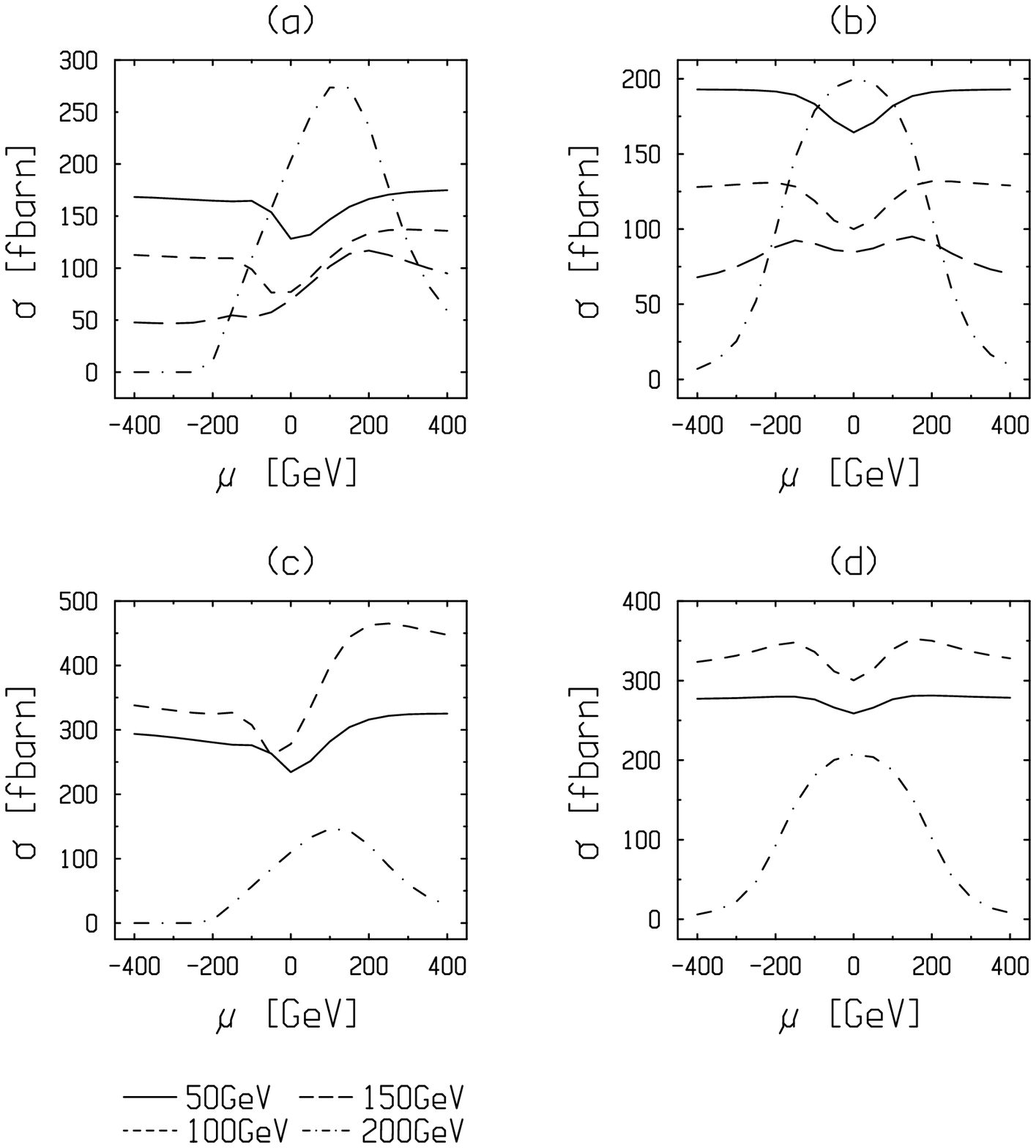}
\end{center}
\caption{\footnotesize  \it
The integrated cross sections for the process, $l_J^+l_J^- \to \tchi_1^- l_m^+$,
at a center of mass energy of $200GeV$, are shown as a function of $\mu$ for discrete
 choices of the remaining parameters: (a) 
$\tan \beta =2, \ m_0=50GeV$, (b) $\tan \beta =50, \ m_0=50GeV$, 
(c) $\tan \beta =2, \ m_0=150GeV$ and  (d) $\tan \beta =50, \ m_0=150GeV$,
with $\l_{mJJ}=0.05$. The windows conventions are such that $\tan \beta=2,50$
horizontally and $m_0=50,150GeV$ vertically.
The different curves refer to the values of $M_2$ of $50GeV$
(continuous line), $100GeV$ (dot-dashed line), $150GeV$ (dashed line) and $200GeV$
 (dotted line), as indicated  at the bottom of the figure.  
\rm \normalsize }
\label{chsect1}
\end{figure}

\begin{figure}
\begin{center}
\leavevmode
\psfig{figure=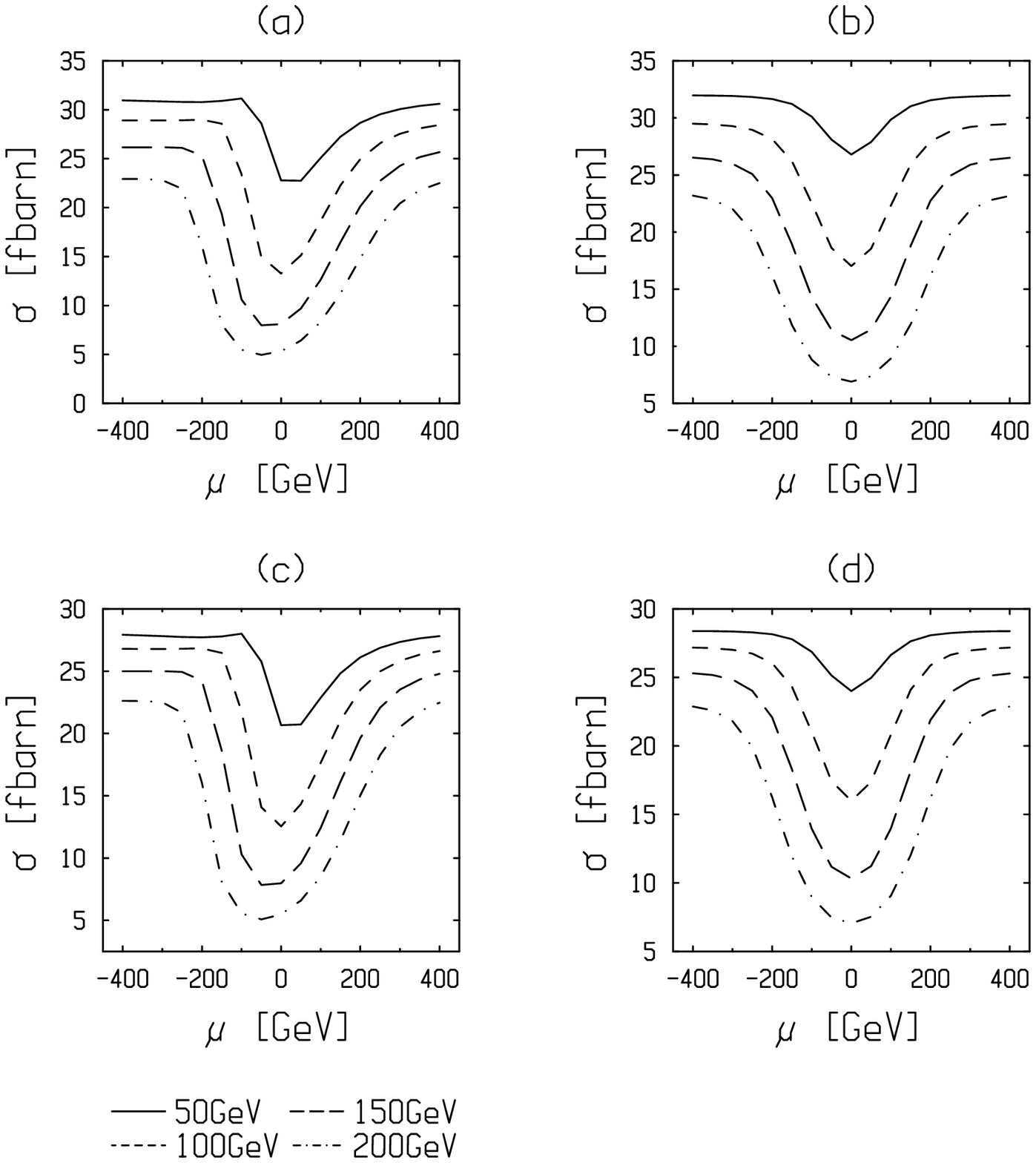}
\end{center}
\caption{\footnotesize  \it
The integrated cross sections for the process, $l_J^+l_J^- \to \tchi_1^- l_m^+$,
at a center of mass energy of $500GeV$, are shown as a function of $\mu$ for discrete
 choices of the remaining parameters: (a) 
$\tan \beta =2, \ m_0=50GeV$, (b) $\tan \beta =50, \ m_0=50GeV$, 
(c) $\tan \beta =2, \ m_0=150GeV$ and  (d) $\tan \beta =50, \ m_0=150GeV$,
with $\l_{mJJ}=0.05$. The windows conventions are such that $\tan \beta=2,50$
horizontally and $m_0=50,150GeV$ vertically.
The different curves refer to the values of $M_2$ of $50GeV$
(continuous line), $100GeV$ (dot-dashed line), $150GeV$ (dashed line) and $200GeV$
 (dotted line), as indicated  at the bottom of the figure.  
\rm \normalsize }
\label{chsect2}
\end{figure}

\begin{figure}
\begin{center}
\leavevmode
\psfig{figure=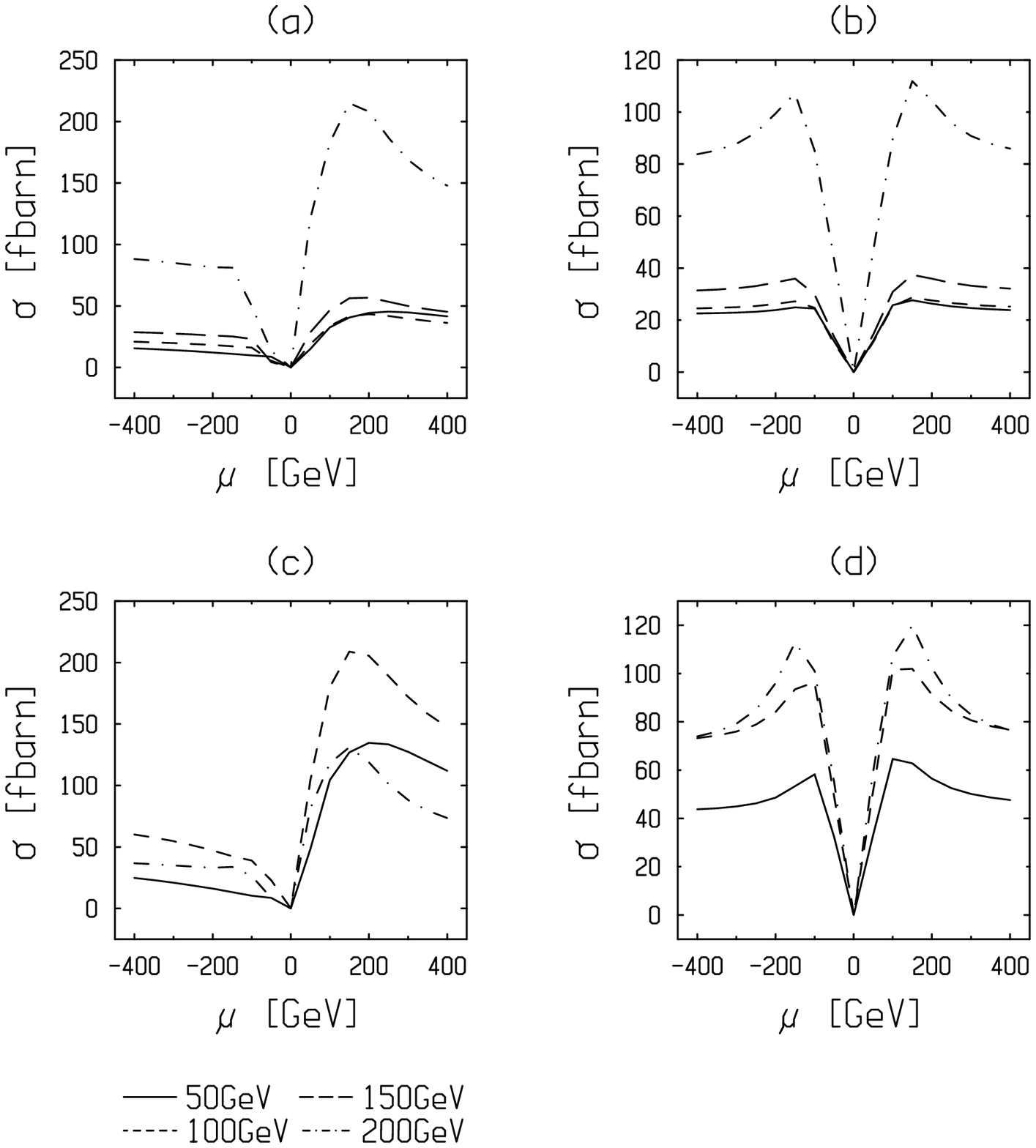}
\end{center}
\caption{\footnotesize  \it 
The integrated cross sections for the process, $l_J^+l_J^- \to \tchi_1^0 \bar \nu_m$,
at a center of mass energy of $200GeV$, are shown as a function of $\mu$ for discrete
 choices of the remaining parameters: (a) 
$\tan \beta =2, \ m_0=50GeV$, (b) $\tan \beta =50, \ m_0=50GeV$, 
(c) $\tan \beta =2, \ m_0=150GeV$ and  (d) $\tan \beta =50, \ m_0=150GeV$,
with $\l_{mJJ}=0.05$. The windows conventions are such that $\tan \beta=2,50$
horizontally and $m_0=50,150GeV$ vertically.
The different curves refer to the values of $M_2$ of $50GeV$
(continuous line), $100GeV$ (dot-dashed line), $150GeV$ (dashed line) and $200GeV$
 (dotted line), as indicated at the bottom of the figure.
\rm \normalsize}
\label{neusect1}
\end{figure}

\begin{figure}
\begin{center}
\leavevmode
\psfig{figure=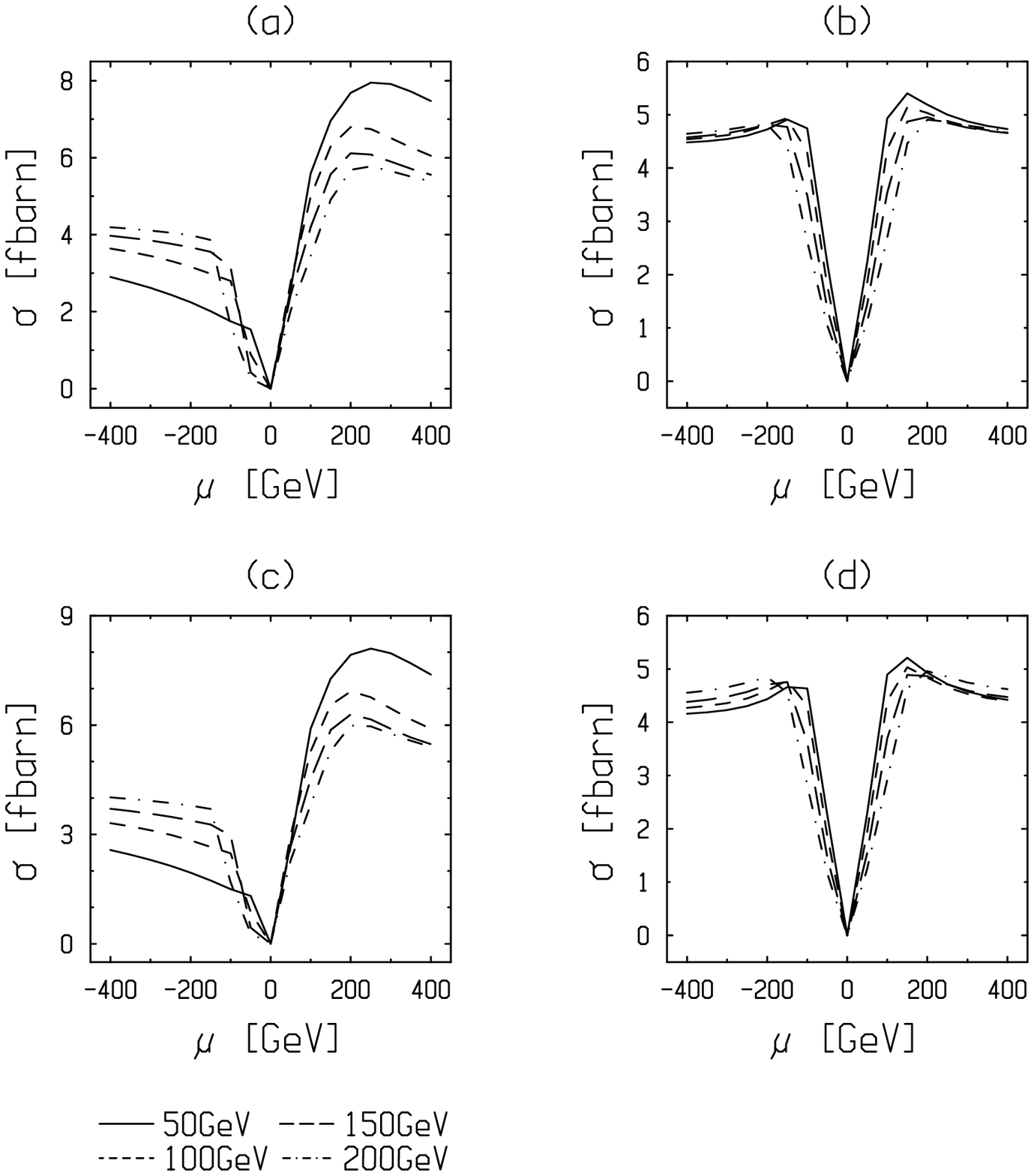}
\end{center}
\caption{\footnotesize  \it 
The integrated cross sections for the process, $l_J^+l_J^- \to \tchi_1^0 \bar \nu_m$,
at a center of mass energy of $500GeV$, are shown as a function of $\mu$ for discrete
 choices of the remaining parameters: (a) 
$\tan \beta =2, \ m_0=50GeV$, (b) $\tan \beta =50, \ m_0=50GeV$, 
(c) $\tan \beta =2, \ m_0=150GeV$ and  (d) $\tan \beta =50, \ m_0=150GeV$,
with $\l_{mJJ}=0.05$. The windows conventions are such that $\tan \beta=2,50$
horizontally and $m_0=50,150GeV$ vertically.
The different curves refer to the values of $M_2$ of $50GeV$
(continuous line), $100GeV$ (dot-dashed line), $150GeV$ (dashed line) and $200GeV$
 (dotted line), as indicated at the bottom of the figure.
\rm \normalsize}
\label{neusect2}
\end{figure}

\begin{figure}
\begin{center}
\leavevmode
\psfig{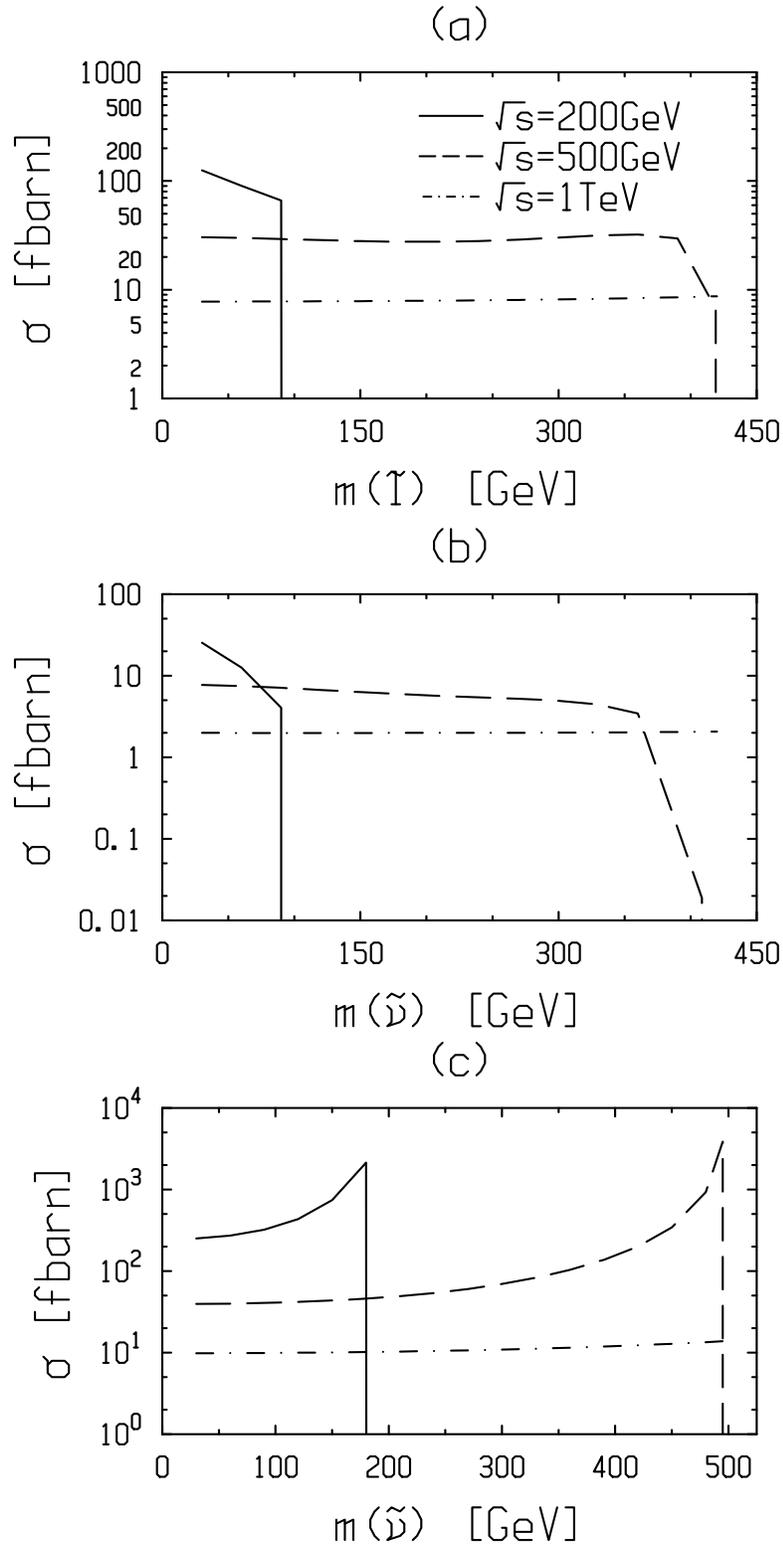}
\end{center}
\caption{\footnotesize  \it
The cross sections for the processes, $l_J^+l_J^- \to \tilde l_m^- W^+$ (a),
$l_J^+l_J^- \to \tilde \nu_m Z^0$ (b) and $l_J^+l_J^-
\to \tilde {\nu}_m  \ \gamma$ (c), 
are shown as a function of the slepton mass and the sneutrino mass,
for $\l_{mJJ}=0.05$. The three values of the center of mass energies 
considered are $200, \ 500$ and $1000GeV$, as quoted in the top window.
\rm \normalsize}
\label{slsect}
\end{figure}

The total production rates are evaluated by taking the angular integral,
$\s=\int_{-x_m}^{+x_m} {d \s \over dx} dx, \ [x=\cos \theta]$,
over the differential cross sections which are given explicitly in
 eqs.(\ref{eqrc4a})-(\ref{eqrc4e}) in Appendix \ref{seca}.
 To follow the usual practice we shall set an angular cut-off to account for 
 the poor detection condition along the beam pipe: $170^o > \theta_m > 10^o  $,
 corresponding to $x_m =\cos \theta_m = 0.9848$.

\subsubsection{Inos production}

The results for the integrated rates of the production of the lowest mass eigenstates
 $\tchi_1^-$ and $\tchi_1^0$, 
 at LEPII energies, are displayed in Figures \ref{chsect1} and \ref{neusect1}, respectively. The inos production rates 
depend smoothly on $\tan \beta$, and on the mass parameters, $\mu, \ m_0, \ M_2$, in a way 
which closely reflects on the mass spectrum. Thus, the symmetry under $\mu \leftrightarrow -\mu$
 is upset only for low $\tan \beta$ and the rates decrease with increasing $M_2$. The only
cases where fast variations of rates arise are 
 for values of $m_0$ and $M_2$ at which the center of mass energy hits on 
the sneutrino s-channel pole, $\sqrt s=m_{\tilde \nu}$. As $m_0$ increases, the resonance occurs
 at smaller values of $M_2$ since the sneutrino mass depends on $ M_2, \ m_0 $ and $ \tan \beta$
 (see eq.(\ref{specsugra})). The pole cross sections themselves, as parametrized by the 
conventional formula, 

\begin{eqnarray}
\s (l^+l^- \to X)&=&{8 \pi s \over m_{\tilde \nu}^2}{\G(\tilde \nu \to l^+l^-) 
\G(\tilde \nu \to X) \over (s-m_{\tilde \nu}^2)^2+\G_{\tilde \nu}^2} \cr
& \approx & 4 \ 10^8 ({100GeV\over m_{\tilde \nu}})^2 B(\tilde \nu \to l^+l^-)
B(\tilde \nu \to X) \ fbarns,
\label{sigmapole}
\end{eqnarray}

can grow to values several order of magnitudes higher. For clarity, 
we have refrained from 
drawing the cross sections close to the resonant energy in the same plot. This is the reason
why  the curves corresponding to $M_2=150GeV$ do not appear in Figures \ref{chsect1}(c)(d) and
 \ref{neusect1}(c)(d). The effect of the pole can be seen for $M_2=200GeV$ in Figures
\ref{chsect1}(a)(b) and \ref{neusect1}(a)(b). We note also that for 
$\mu=0$, $\tchi^0_1$ is a pure higgsino and the $\tchi^0_1$ production cross section vanishes.
The results for inos production rates at NLC or $\mu^+ \mu^- $ colliders center of mass energies are displayed in Figures \ref{chsect2} and \ref{neusect2}. 
The drop with respect to the LEPII energies is nearly by one order of magnitude.
The second neutralino production rates, $\sigma(\tchi_2^0)=
\sigma(l^+_J l^-_J \to \tchi_2^0 \nu_m)$, when this is kinematically allowed, turns out
 to be of the same order of magnitude as $\sigma(\tchi_1^0)$. For $\sqrt s=500GeV$,
$\sigma(\tchi_1^0)$ and $\sigma(\tchi_2^0)$ are numerically close throughout the parameter
 space of our model. However, for $\sqrt s=200GeV$, there are regions (large $\tan \beta$,
 $\mu<0$) where one has $\sigma(\tchi_2^0) \approx 2 \sigma(\tchi_1^0)$ and other regions
 (low $\tan \beta$, $\mu>0$) where one rather has $\sigma(\tchi_2^0) \approx {1 \over 2}
 \sigma(\tchi_1^0)$. As for the production rate of the second chargino,
 $\sigma(\tchi_2^-)$, this is always nearly an order of magnitude below
 $\sigma(\tchi_1^-)$.

\subsubsection{Sleptons production}

The slepton and sneutrino production rates depend solely on the sleptons masses and $\l_{mJJ}$.
 The results, obtained by setting $m_{\tilde l}=m_{\tilde \nu}$, are displayed 
in Figure \ref{slsect} for three values of the center of mass energies. 
An account of the mass difference between
 $m_{\tilde l}$ and $m_{\tilde \nu }$ would not change the numerical results in 
any significant way.

 The differential cross section
for the reaction $l_J^+l_J^- \to \tilde \nu \gamma$ 
must be treated with special care because
 of its extreme sensitivity at the end points, $x= \pm 1$, in the limit of vanishing
 electron mass, $m_e \to 0$. As appears clearly on the expression of the squared momentum
 transfer variable, $t=(k'-p')^2=m_{\g}^2-{1 \over 2} (s-m_{\tilde \nu}^2+m_{\g}^2)
(1-{k \over E_k}{p \over E_p}x)$, for $m_{\g}=0$, the t-channel amplitude 
has a collinear  singularity, 
$t \to 0$ as $x \to 1$. An analogous  collinear  
singularity occurs for the u-channel amplitude,
 $u=(k-p')^2 \to 0$ as $x \to -1$. Imposing the cut-off on the center of mass 
angle $\t$ makes the regularisation of collinear singularities pointless.

 In the limit of vanishing $m_{\g}$, 
independently of $x$ and $m_e$, the sneutrino production  cross section
becomes infinite at the limiting energy point, $\sqrt s = m_{\tilde \nu}$. 
This accounts for the property of the numerical results for the 
integrated cross section to rise with $m_{\tilde \nu}$, as seen in Figure
 \ref{slsect}(c). However, if one were to set $m_{\g}$
 at, say, the $\rho$-meson mass, in line with the vector meson dominance hypothesis, 
one would rather find the opposite behaviour with respect to the dependence on
 $m_{\tilde \nu}$. Observe that the increase of the cross section with $m_{\tilde \nu}$
corresponds to the fact that, for $m_{\tilde \nu} \approx \sqrt s$, the 
process $l_J^+l_J^- \to \tilde \nu \g$ behaves like a sneutrino resonant production,
 accompagnied by the initial state radiation of a soft photon. 

\subsubsection{Discussion}

In summary, the single production rates range from 
several 10 's of $fbarns$ to a few 100's 
of $fbarns$ at LEP energies and several units to a few 10's of $fbarns$ at NLC energies.
 Therefore, the superpartners single production are at the limit of observability
 for LEPII assuming an 
 integrated luminosity per year of $200pb^{-1}$  at $\sqrt s=200GeV$. The prospects 
for single production should be rather good at 
NLC \cite{Lumi} and $\mu^+ \mu^- $ colliders \cite{Lumi2}
 since the assumed integrated luminosity per
year is expected to be about $50fb^{-1}$ at $\sqrt s=500GeV$.
 Moreover, it is important to note here
 that had we considered for the RPV coupling constants, constant values for the product
 $\l_{mJJ}({m_{\tilde l_R}\over 100GeV})$, rather than for $\l_{mJJ}$, the rates would get 
an important amplification factor $({m_{\tilde l_R}\over 100GeV})^2$ for increasing 
superpartners masses. Note that the slepton involved in the  bound is of right chirality
and thus is of opposite chirality than the slepton involved in the rate. Of course, the masses 
of $\tilde l_L$ and $\tilde l_R$ are related in a given model.
At $\sqrt s=500GeV$ and assuming $\l_{ijk} \geq 0.05$, all the single 
production reactions should be potentially observable over a broad region of parameter space.
The slepton production reactions could then probe slepton masses up to $400GeV$
 (Figure \ref{slsect}(a)) and sneutrino masses up to $500GeV$ (Figure \ref{slsect}(c)).
 The ino production reactions could probe 
a large region of the parameter plane $(\mu,M_2)$,
 since the dependence on the parameters $m_0$ and $\tan \beta$ is smooth.
To strengthen our conclusions, it is necessary to examine the 
signatures associated with the final states, which is the subject of the next section.

\subsection{Branching Ratios}

\label{sectionBR}

\begin{figure}
\begin{center}
\leavevmode
\psfig{figure=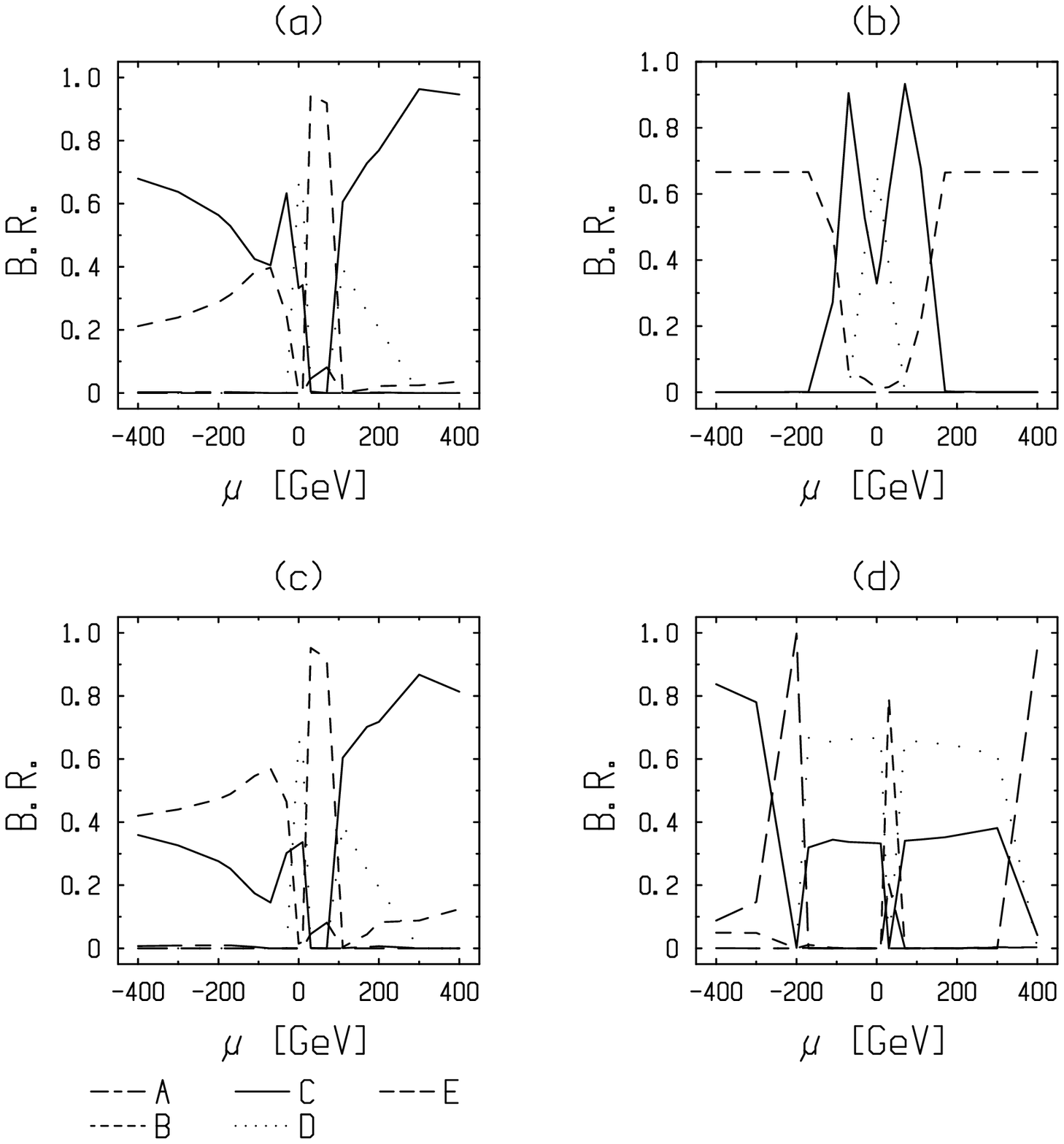}
\end{center}
\caption{\footnotesize  \it
Branching ratios for the chargino $\tchi^-_1$ decays  
as a function of $\mu$. The results in the four windows are obtained 
with the following choices  for the parameters, $[(M_2(GeV),m_0(GeV),\tan 
\beta,\l_{ijk})$, $m_{\tilde \nu_L}(GeV),m_{\tilde l_L}(GeV)]$: 
(a) $[(80,20,2,0.05)$, $53.19,81.66]$,  (b) $[(80,20,50,0.05)$, $34.20,86.97]$, 
(c) $[(80,20,2,0.1)$, $53.19,81.66]$, (d) $[(200,100,2,0.05)$, $195.6,205.2]$.
The final states are labeled by the letters, A,B,C,D,E, which have the same  meaning
as in Table \ref{tabloC}. 
\rm \normalsize }
\label{br1}
\end{figure}

\begin{figure}
\begin{center}
\leavevmode
\psfig{figure=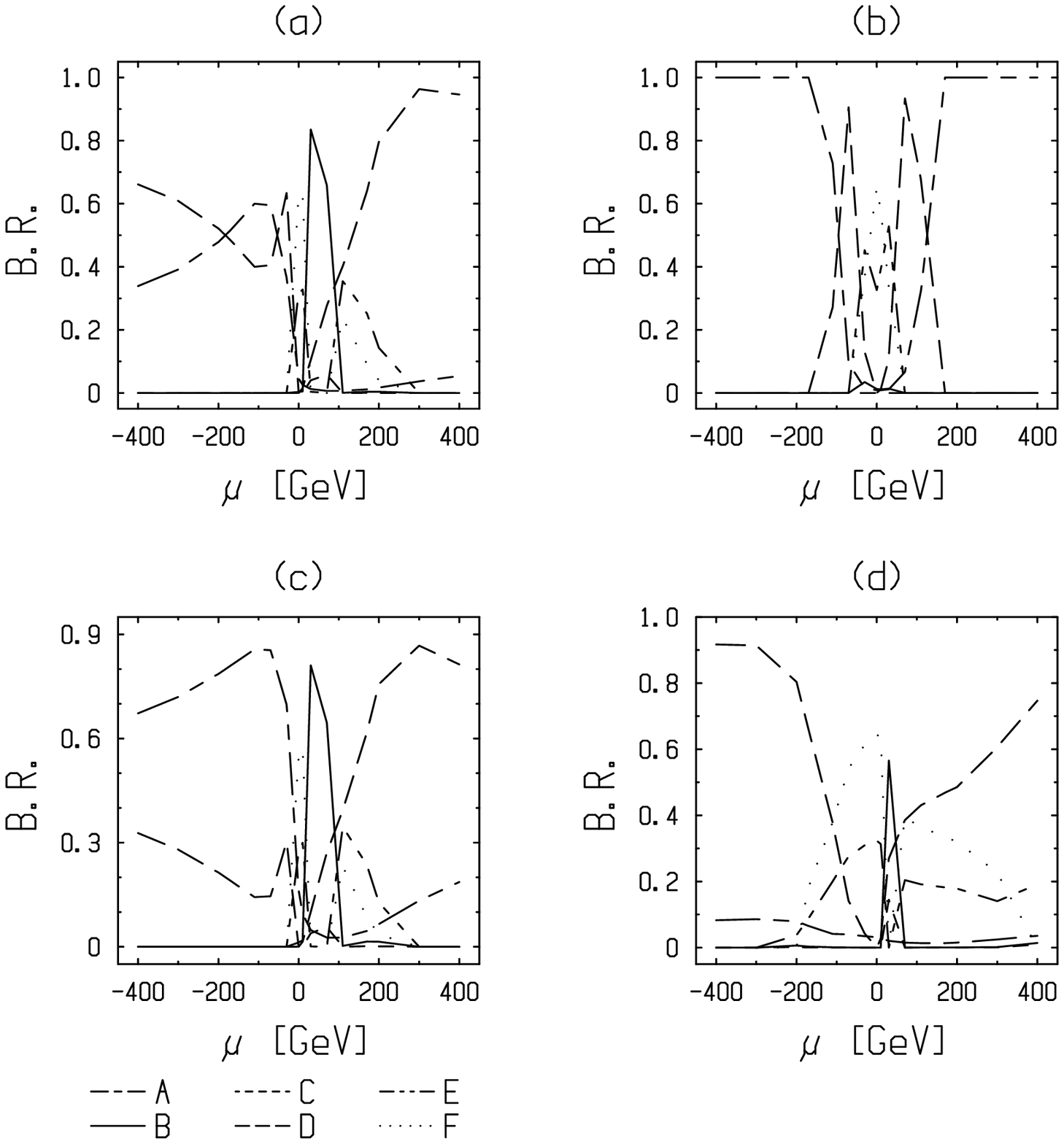}
\end{center}
\caption{\footnotesize  \it
Branching ratios of the sneutrino decays  
as a function of $\mu$. The results in the four windows are obtained 
with the following choices  for the parameters, $[(M_2(GeV),m_0(GeV),\tan 
\beta,\l_{ijk})$, $m_{\tilde \nu_L}(GeV),m_{\tilde l_L}(GeV)]$: 
(a) $[(80,20,2,0.05)$, $53.19,81.66]$,  (b) $[(80,20,50,0.05)$, $34.20,86.97]$, 
(c) $[(80,20,2,0.1)$, $53.19,81.66]$,  (d) $[(200,100,2,0.05)$, $195.6,205.2]$.
The final states are labeled by the letters, A,B,C,D,E,F, which have the same  meaning
as in Table \ref{tabloA}.
\rm \normalsize}
\label{br2}
\end{figure}

\begin{figure}
\begin{center}
\leavevmode
\psfig{figure=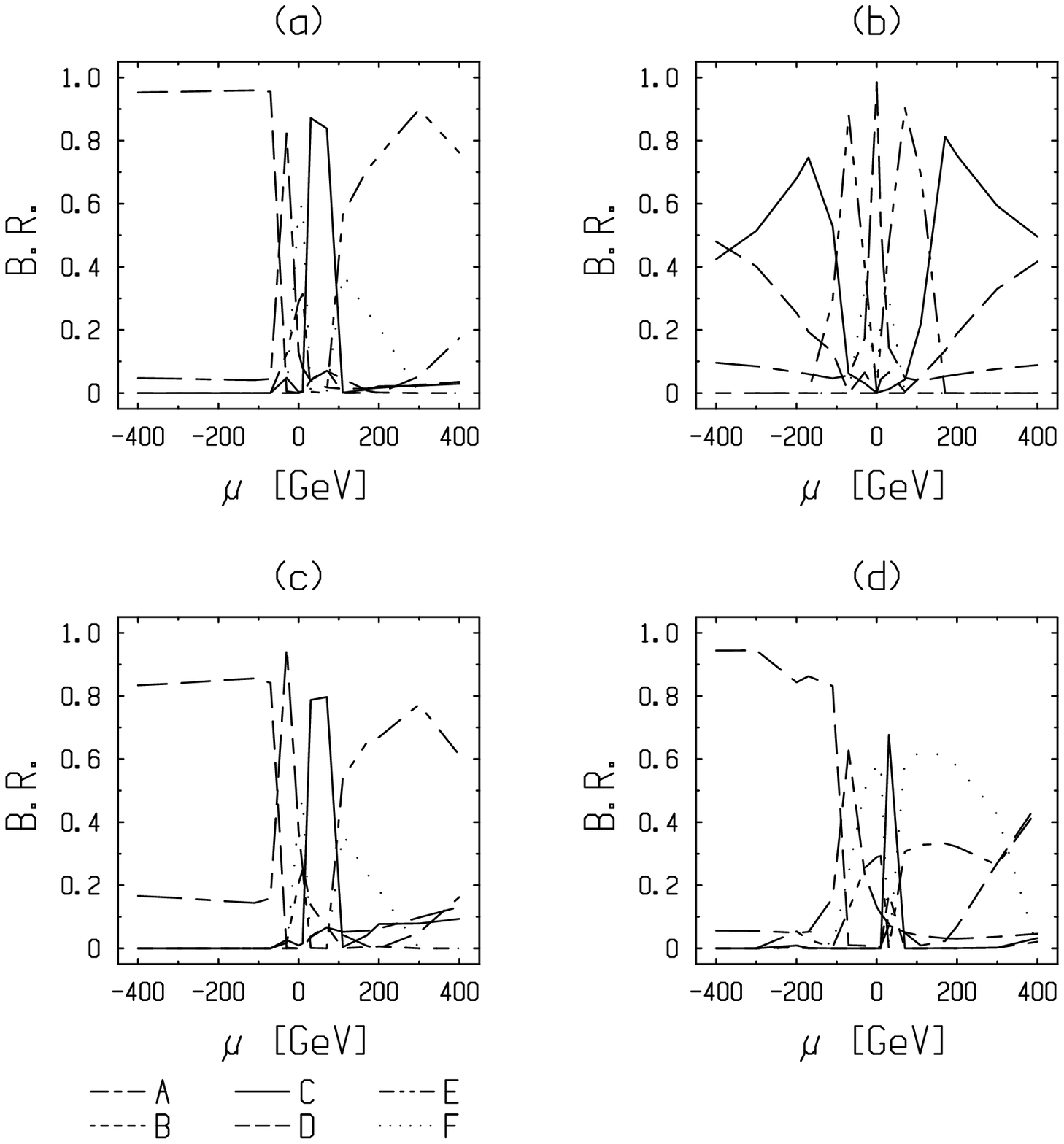}
\end{center}
\caption{\footnotesize  \it
Branching ratios of the slepton decays 
as a function of $\mu$. The results in the four windows are obtained 
with the following choices  for the parameters, $[(M_2(GeV),m_0(GeV),\tan 
\beta,\l_{ijk})$, $m_{\tilde \nu_L}(GeV),m_{\tilde l_L}(GeV)]$:  
(a) $[(80,20,2,0.05)$,  $53.19,81.66]$,  (b) $[(80,20,50,0.05)$,  $34.20,86.97]$, 
(c) $[(80,20,2,0.1)$,  $53.19,81.66]$, (d) $[(200,100,2,0.05)$,  $195.6,205.2]$.
The final states are labeled by the letters, A,B,C,D,E,F, which have the same meaning
as in Table \ref{tabloB}.
\rm \normalsize}
\label{br3}
\end{figure}

\begin{figure}
\begin{center}
\leavevmode
\psfig{figure=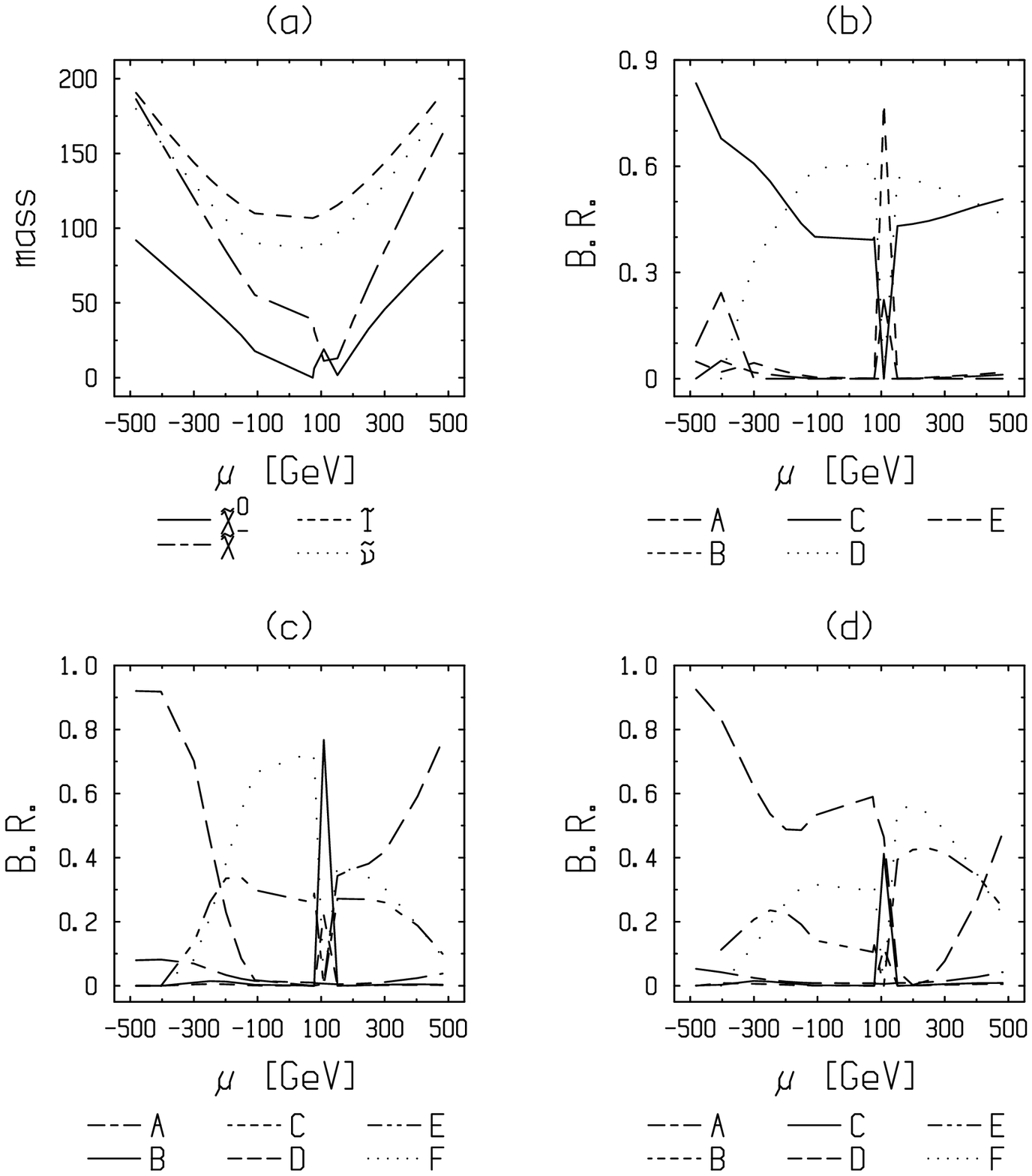}
\end{center}
\caption{\footnotesize  \it
Mass spectrum of the supersymmetric particles (a), in GeV, and branching ratios for the decays
of the chargino (b), sneutrino (c) and slepton (d), as a function of $\mu$.
The results are obtained for $m_0=100GeV$, using equation (\ref{ewsb}).
 The final states in figures (b),(c),(d) are labeled by the letters, A,B,... 
which have the same meaning as in tables \ref{tabloC},\ref{tabloA},\ref{tabloB}, respectively.
\rm \normalsize}
\label{bratewsb}
\end{figure}

In the narrow resonance approximation, the partial transition rates are readily 
obtained by multiplying the total rates for each reaction with the decay branching
fractions. The various final states for each of the $2 \to 2$ single production reactions
have been listed in Tables \ref{tabloC}, \ref{tabloA} and \ref{tabloB}. The leptons
family configurations in the final states will depend on the hypothesis for the
RPV coupling constant (single or pair dominance).

With the purpose of testing characteristic points of the parameter space,
we have evaluated the branching ratios for the decays of the superpartners, namely,
$\tchi_1^{\pm},\tilde l^{\pm}$ and $\tilde \nu$,
for variable $\mu$ at discrete choices of $M_2$, $m_0$ and $\tan \beta$, such that 
the main typical cases in the ordering of the masses $m_{\tchi_1^0}$, $m_{\tchi_1^{\pm}}$,
$m_{\tilde l}$, $m_{\tilde \nu}$, can be explored. The results are shown in Figures
\ref{br1}, \ref{br2} and \ref{br3}, for the chargino, the
sneutrino and the slepton decays, respectively. The curves for the various branching ratios
are distinguished by the same letters (numbers) as those used in Tables \ref{tabloC}, 
\ref{tabloA} and \ref{tabloB} to label the various final states (decay processes).
We shall now discuss in turn the various superpartner decay schemes corresponding
 to the five single production reactions.

\subsubsection{Lowest mass Neutralino}
\label{neutralinosection}

The branching ratios for the $\tchi_1^0$ desintegrations are best analysed separately.
For convenience, we do not treat the cases, $m_{\tchi_1^0}>m_{\tilde q}$ 
and  $m_{\tchi_1^0}>m_{\tchi_1^{\pm}}$, since these arise marginally 
in most of the currently favored models (supergravity or gauge mediated soft supersymmetry breaking).
The cascade decays which occur if $m_{\tchi_1^{\pm}}<m_{\tchi^0_1}$ are also not considered
since the corresponding region of the parameter space (Figure \ref{spec}(c)) is 
forbidden by the experimental constraints on the inos masses. Thus,
the process $l_J^+l_J^- \to \tchi_1^0 \nu_m$ will only generate events with 2 leptons +
 $\Eslash $. At this point, it is necessary to specialize our discussion to a single 
dominant coupling constant hypothesis, assuming $\l_{ijk} \neq 0$ not necessarily
 identical to $\l_{mJJ}$. One may distinguish the following four distinct cases.
For an LSP $\tchi_1^0$, namely, $m_{\tchi_1^0}<m_{\tilde l}, 
m_{\tilde \nu}$ (Case 1), only the direct RPV three-body decays, $\tchi_1^0 \to
\bar \nu_i \bar l_j  l_k$, $\tchi_1^0 \to \nu_i l_j  \bar l_k$, are allowed.
The branching ratios are then determined on the basis of simple combinatoric arguments.
For a dominant coupling constant, say, $\l_{m11}$, there are four final states:
 $\nu_1 l^-_m e^+$, $\bar \nu_1 l^+_m e^-$, $\nu_m e^- e^+$ and $\bar \nu_m e^+ e^-$. 
Accordingly, the branching ratios of $\tchi_1^0$ into two charged leptons  will depend on the type
(flavor,charge) of the final state: The branching ratios equal ${1 \over 2}$ for the flavor diagonal
$e^+e^-$ or flavor non diagonal $l^{\pm}e^{\mp}$ channels, ${1 \over 4}$ for the
fixed charges and flavors $l^+e^-$ or $l^-e^+$ channels and $1$ for the lepton-antilepton
 pairs of unspecified flavors.
For a dominant coupling constant $\l_{ijk} \neq \l_{m11}$, an analogous result is obtained.
For $m_{\tchi_1^0}>m_{\tilde l},m_{\tilde \nu}$ (Case 2), the branching ratio for $\tchi_1^0$ decay is,
\begin{eqnarray}
B(\tchi_1^0 \to \bar \nu_i \bar l_j  l_k)={\G (\tchi_1^0 \to \tilde l_j \bar l_j)
B(\tilde l_j \to \bar \nu_i l_k) + \G(\tchi_1^0 \to \tilde \nu_i \bar \nu_i) 
B(\tilde \nu_i \to \bar l_j l_k) 
\over
3\G (\tchi_1^0 \to \tilde l_j l_j) +3\G (\tchi_1^0 \to \tilde \nu_i \nu_i )}=
{1 \over 3},
\label{eqap3}
\end{eqnarray}
where we have used the fact that in the present case, assuming a dominant coupling constant $\l_{ijk}$,
 $B(\tilde l_j \to \bar \nu_i l_k)=B(\tilde \nu_i \to \bar l_j l_k)=1$. The factors $3$ 
in the denominator account for the number of families. For the 
intermediate case, $m_{\tilde \nu}>m_{\tchi_1^0}>m_{\tilde l}$ (Case 3), there occur 
contributions from 2-body RPC decays and 3-body RPV decays, such that:
\begin{eqnarray} 
B(\tchi_1^0 \to \bar \nu_i \bar l_j  l_k)={\G' (\tchi_1^0 \to \bar \nu_i \bar l_j l_k )
+ \G (\tchi_1^0 \to \tilde l_j \bar l_j)
B(\tilde l_j \to \bar \nu_i l_k) 
\over
\G' (\tchi_1^0 \to \bar \nu_i \bar l_j l_k ) + 3\G (\tchi_1^0 \to \tilde l_j l_j) } \simeq
{1 \over 3},
\label{eqap4}
\end{eqnarray} 
where the prime on $\G'$ is a reminder to indicate that the decay width includes only the 
contribution from a virtual sneutrino exchange. The approximate equality in eq.(\ref{eqap4})
 derives from the fact that $\G' (\tchi_1^0 \to \nu_i  \bar l_j l_k)<<\G 
(\tchi_1^0 \to \tilde l_j \bar l_j)$, based on the expectation that an RPV 3-body 
decay should be much smaller than an RPC 2-body decay. An analogous argument to 
that of case 3 holds 
for the other intermediate case, $m_{\tilde l}>m_{\tchi_1^0}>m_{\tilde \nu}$ (Case 4).
For the cases 2, 3 and 4, the multiplicity factors are the same as for the case 1.
The $\tchi_1^0$ process may occur at the end stage in the decays of $\tchi_1^-$, $\tilde l$ 
and $\tilde \nu,$ to be discussed below. The associated
$\tchi_1^0$ decay multiplicity factors
 for the two leptons final states will then take the same values 
 as quoted above for the various 
selection criteria. In quoting numerical results below, we shall, for convenience, 
assume the case of unspecified lepton flavor and charge and thus will set the
 multiplicity factors to unity.

\subsubsection{Lowest mass Chargino}
The results  in Fig. \ref{br1} for the  high $\tan \beta $ case 
show a high degree of symmetry with respect to  $\mu \leftrightarrow -\mu $,
which arises from the symmetry in
 the inos mass spectrum (Figure \ref{spec}(a)(b)).  
As can be seen from Figure \ref{br1}(a), a dominant mode for the chargino 
at high values of $\vert \mu \vert$ is the cascade decay, 
$\tchi^- \to \tchi^0 l^- \bar \nu$, since this  occurs via the two-body decay,
$\tchi^- \to l^- \tilde {\bar \nu}$ (C(7)). Indeed, for these high values of $\mu$,
one has $m_{\tilde {\bar \nu }}<m_{\tchi^-}$. This two-body decay competes with 
the other two-body decay E, $\tchi^- \to \tchi^0 W^-$, when the latter 
is kinematically allowed,
as is the case for $\mu < -200GeV$ in Figure \ref{br1}(d). The difference between the
values of the branching ratios C(7) and E is explained by the relative phase space 
factors of the associated rates. The RPV direct decays (A and B) are three-body decays
with small coupling constant and are thus suppressed. In the case, $m_{\tilde 
{\bar \nu}}<m_{\tchi^0}<m_{\tchi^{\pm}}$, the only open channel for the sneutrino is,
$\tilde {\bar \nu} \to l \bar l$, so that the dominant mode for the chargino decay
is the RPV decay B(4) (high values of $\vert \mu \vert$ in Figure \ref{br1}(b)).
Even for $m_{\tilde {\bar \nu}} \approx m_{\tchi^0}$, the channel B(4) is competitive
due to a small phase space ($\mu \simeq -100GeV$ in Figure \ref{br1}(a)). In this case, 
for $\l_{ijk}=0.1$, the direct RPV decay B(4) can become dominant (moderate negative 
values of $\mu$ in Figure \ref{br1}(c)). For small values 
of $\vert \mu \vert$, the difference between the two dominant leptonic (C) 
and hadronic (D) cascade decays is due to the flavor and 
color factors. We note also that in a small interval of $\mu$ near $\mu=0$,
$m_{\tchi^0_1}>m_{\tchi^{\pm}_1}$ (see Figure \ref{spec}), and consequently the 
only open channels are the direct RPV decays (Figure \ref{br1}(a)(c)(d)). 
In this region, the direct RPV decay A(1) is
negligible  because the branching ratio depends on $U_{11}$ which is small \cite{Pol}.
 In conclusion,
the highest branching ratios are associated with the cascade decays, C,D and E, except
 for the case
in which the sneutrino is the LSP, where they are associated with the RPV decays B. 
The range of $\mu$ for which the chargino $\tchi_1^-$ is the LSP is excluded by the 
experimental constraints on the inos masses  (see Figure \ref{spec}).

\subsubsection{Sneutrino}
We turn now our attention to the sneutrino decays. 
For high values of $\mu$, the cascade decay D has the highest probability
(Figure \ref{br2}(a)(d)) since the decay into chargino is either kinematically forbidden
or suppressed by a small phase space.
As for the chargino study, the RPV direct decay A is of course small except when 
the competitive channel is reduced by a small phase space factor ($\mu \approx -100GeV$ 
in Figure \ref{br2}(a)). In such a case, the RPV direct decay A may be important for values 
of $\l_{ijk}$ near $0.1$ (negative $\mu$ in Figure \ref{br2}(c)). 
 When the sneutrino is the LSP,
the RPV direct decay has a branching ratio equal to unity (Figure \ref{br2}(b)).
 For small $\vert \mu \vert$,
the decays E(6) and F(8) through charginos dominate the decay D through neutralinos.
 The reason is that for $\mu=0$, $\tchi^0_1$ is a pure higgsino, whose couplings are weak. 
In the so called higgsino limit, $\mu \to 0$ \cite{Pol}, the decays B and C(3)
 are small since they occur through the
$\tchi^-$ RPV direct decays. However, they have the highest probability if, $m_{\tchi^0_1}>
m_{\tchi^{\pm}_1}$ (Figure \ref{br2}(a)(c)(d)). The relation between the leptonic (E) and 
hadronic (F) cascade decays
can be explained in the light of the study on the chargino. We conclude that the 
cascade decays, B, D, E and F, are always the dominant modes, except when the 
sneutrino is the LSP.

\subsubsection{Slepton}

Finally, we concentrate on slepton decays. For high values of $\mu$, the cascade decays 
via charginos are reduced because of a small phase space ($\vert \mu \vert \approx 400GeV$ 
in Figure \ref{br3}(b) or for $\mu<0$ in Figures \ref{br3}(c)(d)) or even closed 
(for $\mu<0$ in Figure \ref{br3}(a)). In these cases,
the decay D via neutralinos dominates. Elsewhere, the decays via charginos have higher 
branching ratios (for $\mu>0$ in Figures \ref{br3}(a)(d))
since larger coupling constants are  involved. 
In the higgsino limit, the slepton cascade decay D via $\tchi^0_1$ is suppressed for the same 
reason as in the sneutrino study. 
The decay via $\tchi^-_1$ is then dominating. The interpretation of the difference
between the decays, B, C, E and F, via charginos is based on the specific behaviours of the chargino
 branching ratios which have already been described above. We see in Figure \ref{br3}(c), that for 
$\l_{ijk}=0.1$, the RPV direct decay A is still very reduced.
 This is due to the important phase space
for the slepton decay into neutralino. Lastly, a new phenomenon appears for the slepton case.
In the higgsino limit ($\mu \to 0$) at large $\tan \beta$, the matrix element $U_{11} \to 0$
which forces the vertex $\tilde l \ \tchi^{\pm} \ \nu $ (see eq.\ref{eqpr6} in Appendix
 \ref{secappa}) and the branching ratios for the cascade decays through the chargino to vanish \cite{Pol}.
 This is the explanation of the fact that for $\mu \simeq 0$, one observes a peak of the 
direct RPV decay branching fraction (Figure \ref{br3}(b)). Similar peaks are also observed
 at shifted $\mu <0$ for the low $\tan \beta$ cases (Figures \ref{br3}(a)(c)(d)). 
However this behaviour appears for ranges of the parameters which are 
forbidden by the bounds on the inos masses (Figure \ref{spec}).
 The conclusion is that the cascade decays have always the highest probability for the reason 
that the L-chirality slepton cannot be the LSP in generic supergravity models.

\subsubsection{Discussion}
In summary, we have learned that the general behaviour of branching ratios is mainly
determinated by the phase space and thus by the ordering of the supersymmetric particles 
masses. We have explored all the characteristic cases,
$m_{\tilde \nu}>m_{\tchi_1^-}>m_{\tchi_1^0}$,
$m_{\tchi_1^-}>m_{\tilde \nu}>m_{\tchi_1^0}$ and  $m_{\tchi_1^-}>m_{\tchi_1^0}>m_{\tilde \nu}$.
For high values of $m_0 $ lying above $M_2$, the sleptons would have masses greater
than the inos masses. We have not analysed this case since
one has then the same situation 
in the mass ordering as for the case of small values of
$\vert \mu \vert$ (except for large
enough values of $m_0$ where the on-shell $W^{\pm}$ production
can take place in $\tilde l$
and $\tilde \nu$ decays G). In this situation, as we have explained
above, the charginos principally decay into neutralinos,
while the sleptons and sneutrinos decay
into charginos. The main conclusion is that the cascade decays are the dominant modes except 
if the sneutrino is the LSP. In this case, the RPV decay, $\tchi_1^- \to l_i \tilde {\bar \nu}_i 
\to l_i l_j \bar l_k$, is dominant for the chargino decays, and the only open channel for 
the sneutrino is of course the direct RPV decay. Besides, for values of $\l_{ijk}$ higher 
than $0.05$, the RPV direct decay branching ratios can reach significant levels for the case where the 
cascade decays are suppressed due to small phase space factors.

The excitation of the second neutralino $\tchi_2^0$ deserves some attention since this may
 have in certain regions of the parameter space comparable, if not larger, production 
rates than 
the excitation of $\tchi_1^0$. Assuming that the direct RPV widths are small enough so 
that the decay chain is initiated by the RPC contributions, then the desintegration mode,
 $\tchi^0_2 \to (\tchi_1^0+l^+ l^-),(\tchi_1^0+\bar \nu \nu)$, will also yield 
2l+$\Eslash$ and 4l+$\Eslash$ final states, respectively, and the other
 desintegration modes, $\tchi^0_2 \to (\tchi_1^+ +l^- \bar \nu,\tchi_1^- +l^+ \nu),
(\tilde {\bar \nu} \nu,\tilde \nu \bar \nu),(\tilde l^{\pm} l^{\mp})$, will yield 
2l+$\Eslash$ and 4l+$\Eslash$ final states according to decay schemes similar to those
 given in Tables \ref{tabloA},\ref{tabloB},\ref{tabloC}. In our supergravity models,
 the $\tchi^0_2$ decay into $\tchi^{\pm}_1$ should be suppressed by a small phase space
 (Fig.\ref{spec}). To determine which of the decay modes, $\tchi^0_2 \to \tchi^0_1, \
 \tilde l$ or $ \tilde \nu$, leads to the dominant signal would require a detailed
 comparison of \brs at the initial as well as the subsequent stages.

Let us ask in what way would alternate hypotheses 
on the family dependence affect our conclusions.
 Especially regarding the multiplicities of final states, this is relevant for the cases,
$m_{\tchi_1^-}>m_{\tilde l},m_{\tilde \nu}$, where the chargino can  
cascade decay to on-shell
 sleptons or sneutrinos (A(2) and B(4) in Table \ref{tabloC}).
As we have emphasized in  the last paragraph of 
Section \ref{sectionD}, the chargino decays have a multiplicity of 2 
for three degenerate families of sleptons. 
For the case of two degenerate families, labeled by the indices, $m,n$, assuming a dominant RPV
coupling constant $\l_{ijk}$, the multiplicity equals 2 for $(m,n)=(i,j)$, 
since the two sleptons from families $i$ and $j$ 
can be produced on-shell, and equals  1 for $m=k$ or $n=k$.
For the physically interesting case of a single low mass family, labeled by the index, $m$,
one finds that the multiplicity equals 1 for $m \neq k$ and 0 otherwise. 
The conclusion is that the RPV contributions A and B (in Table \ref{tabloC}) to 
the chargino branching ratios increase as the number of slepton families, which are lower in mass 
than the chargino, becomes higher. 
This effect, which is quite small, would affect the branching ratios in parameters regions for which the RPV
contributions A and B are not weak, that is for $\mu<0$ in Figures \ref{br1}(a)(b)(c).

In Figure \ref{bratewsb}, we present results for the branching fractions for fixed $m_0$ in the 
infrared fixed point model with electroweak symmetry breaking. 
In this constrained version, 
where $m_{1/2}$ varies with $\mu$, the 
dependence on $\mu$ is rather similar to that of the non minimal 
model where we worked  instead with fixed
 $m_0$ and $m_{1/2}$.  However, as we see from the mass spectrum, here
the LSP is the neutralino $\tchi_1^0$ for all the physical ranges of the parameters.
Due to the large mass difference between the $\tchi_1^0$ LSP and the NLSP (next to LSP), 
the cascade decays are the only dominant modes and the branching ratios for the RPV direct decays are 
very weak.

Let us add a few qualitative remarks on the predictions of
 gauge mediated \susy breaking models. In order for the production rates in
 the minimal model \cite{Dimo97} to have the same order
 of magnitude as those obtained in the
 supergravity model of section \ref{sectionR}, one needs a parameter $\Lambda
 ={F\over M} \simeq 10^{4}GeV$, using familiar notations for the \susy
 breaking scale ($\sqrt F$) and messenger scale ($M$). 
Concerning the signals, by
 comparing the mean free paths for
 $\tchi^0_1$ (favourite candidate for LSP) in both models, one finds that the
 decay channel to the gravitino, $\tchi_1^0 \to \gamma \ \tilde G$,  becomes competitive
 with the RPV decay channel, $\tchi_1^0 \to  \nu l \bar l$,   for,
 $ {\sqrt {<F>} \over 100TeV} \leq { 10^{-2} \over \sqrt \l }$.

Let us also comment briefly on some of the experimental issues.
A given final state can possibly arise
 simultaneously from several of the single production processes.
 The important 4l+$\Eslash $ 
signal 
which occurs for $\tchi^{\pm},\tilde l^{\pm},\tilde \nu$ productions is one 
such example where one 
may be forced to add all three types of cross sections in comparing with some given
experimental data sample.
Similarly, for most signals, one must typically add the contributions 
from the two charge conjugate partner processes.
Concerning the competition with the standard model background, 
one expects that the most important contributions
to the final states, 2l+$\Eslash $ and 4l, will arise  from the reactions,  $l_J^+l_J^- \to
W^+l^-\bar \nu , \  W^-l^+\nu , \   W^+W^-,\   Z^0 l^+l^-,\   Z^0Z^0,\  Z^0 \gamma$. 
In spite of the large standard model rates of order one picobarn 
at $\sqrt s = 500 GeV$ \cite{DESY2},  one should be able to distinguish
the single production signals by exploiting their specific non diagonal
 flavor character (final state B in Table \ref{tabloB} and A in Table \ref{tabloA}). 
The other multileptons final states, 
generated by the cascade decays, 4l+$\Eslash $, 4l+$Z^0$,
 3l+$Z^0$+$W^{\pm}$+$\Eslash $,... have a standard model background which is negligible. 
The potentially large two photons background processes, induced by 
$\gamma \gamma$ photons pairs radiated by the initial
 leptons, can be significantly reduced by imposing suitable
 cuts on the leptons transverse momenta. Finally, we note that the selection by the
RPV single production of identical helicities for the initial state, $l^+_Hl^-_H$, 
can be exploited to discriminate against the minimal supersymetric standard model
and also the standard model, for which the identical helicities configuration only
appears with the t-channel Z-boson exchange.

\section{Dynamical distributions}

\label{sectionCD}

\begin{figure}
\begin{center}
\leavevmode
\psfig{figure=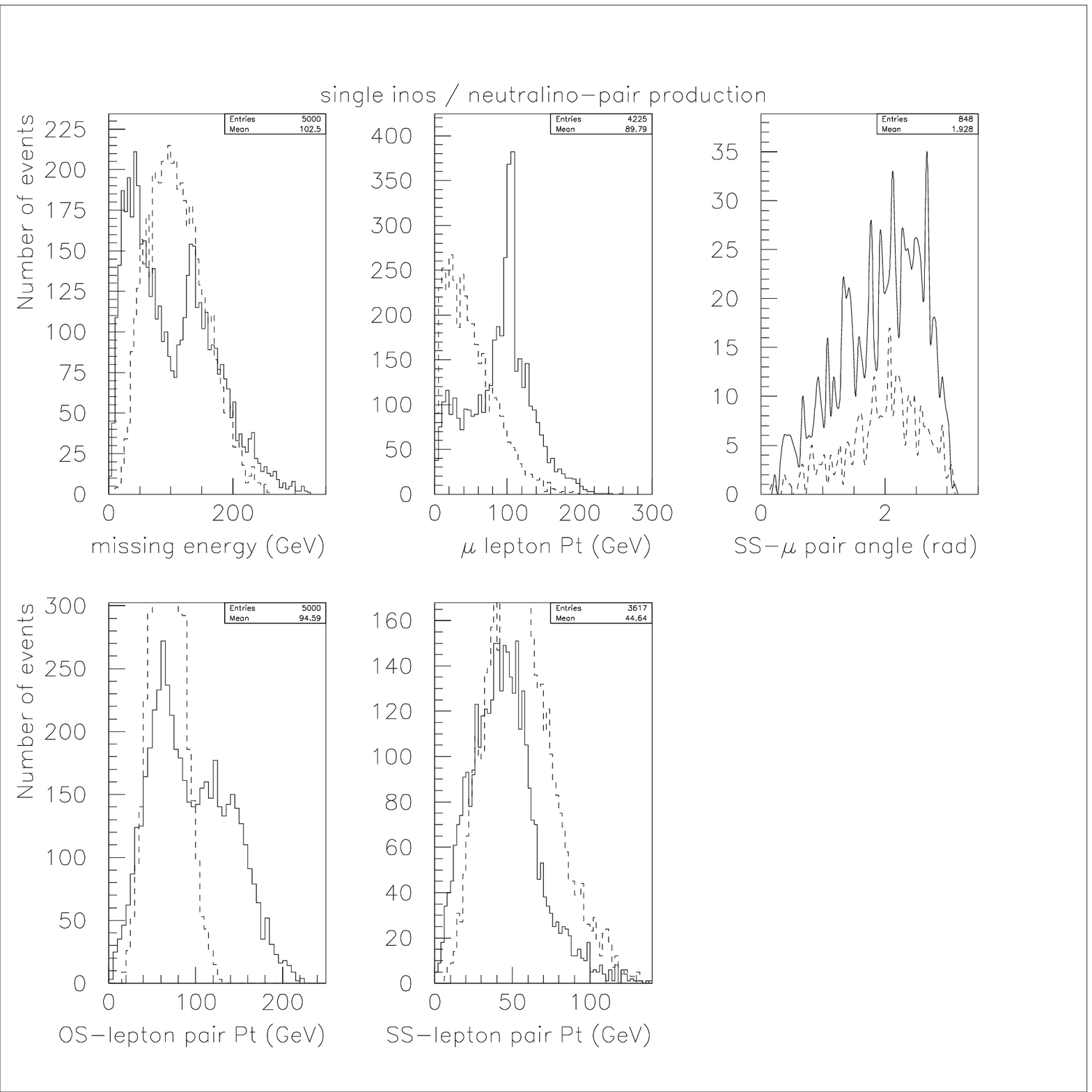}
\end{center}
\caption{\footnotesize  \it
Distributions of missing energy, muon transverse momentum, same sign muon pair angle,
summed transverse momentum for opposite sign (OS) and 
same sign (SS) leptons pairs 
(electrons and muons) for the single production processes, $l_J^+l_J^- \to \tchi_1^{\pm} \mu^{\mp},
\tchi_1^0 \nu_{\mu},\tchi_1^0 {\bar \nu}_{\mu},\tchi_2^0 \nu_{\mu},
\tchi_2^0 {\bar \nu}_{\mu}$ (solid line), and the pair production process,
 $l_J^+l_J^- \to \tchi_1^0 \tchi_1^0$ (dashed line), at a center of mass energy of $350GeV$. 
The parameters values  are, $M_2=250GeV, \ m_0=70GeV, \ \mu=400GeV, \ \tan \beta =2, \ \l_{211}=0.05$.
Events samples, consisting of 5000 events each, are generated for the inos single production 
and neutralino pair production, respectively. 
\rm \normalsize}
\label{MC}
\end{figure}

The distributions of rates with respect to kinematical variables 
associated with the final states offer
helpful means to characterize the underlying production processes. As an indicative study we shall 
present here some characteristic dynamical distributions obtained for the production reactions, 
$l_J^+l_J^- \to \tchi_1^{\pm}l^{\mp},\tchi^0_1 \nu,\tchi^0_1 \bar \nu,
\tchi^0_2 \nu,\tchi^0_2 \bar \nu$, from a Monte Carlo events simulation 
for which we have used the event generator SUSYGEN \cite{Kats1}.
We concentrate on the final state signals of 2l+$\Eslash $,  4l and 4l+$\Eslash $.
Note that for high values of $\mu$, the final state 4l+$\Eslash $ is the dominant mode for the
 chargino, slepton and sneutrino decays. This signal also receives contributions from the reactions,
 $l_J^+l_J^- \to \tilde \nu Z^0 (\tilde {\bar \nu} Z^0),\tilde \nu \gamma (\tilde {\bar \nu}
 \gamma),\tilde l^{\pm} W^{\mp}$, which however are not included in the simulation. 
The standard model background is expected to be small for the 4l+$\Eslash $ signal. The main background 
from the minimal supersymmetric standard model interactions arises from the neutralino RPC pair production, 
$l_J^+l_J^- \to \tchi^0_1 \tchi^0_1$.  Following the analysis in \cite{FUJII},
we consider an incident energy of $\sqrt s =350GeV$ and use a non minimal supergravity model
 for which we choose the set of parameters, $M_2=250GeV,\mu=400GeV,m_0=70GeV,\tan \beta=2$,
 which yields the spectrum,
 $m_{\tchi_1^0}=118.5GeV,m_{\tchi_2^0}=221.4GeV,m_{\tchi^{\pm}_1}=219.1GeV,
m_{\tilde \nu_L}=225GeV,m_{\tilde l_L}=233GeV,
m_{\tilde l_R}=141GeV$. The integrated rates (ignoring acceptance cuts) are,
 for $\l_{mJJ}=0.05 \ [m=2]$,
 $\sigma (\tchi^+_1 \mu^-)=30.9fb,\sigma (\tchi^0_1 \nu_{\mu})=4.8fb,
\sigma (\tchi^0_2 \nu_{\mu})=12.1fb$ and $\sigma (\tchi^0_1 \tchi^0_1)=238.9fb$.
 We consider the following five dynamical variables for  all types of final states:
Invariant missing energy, 
$E_m= \sum_{i \in \nu} E_{i}$ where the sum is over the neutrinos, 
as appropriate to a broken
 R parity situation; Average per event of the $\mu^{\pm}$
 lepton transverse momentum, $P_t(\mu^{\pm})={\sum_i \vert p_t(\mu^{\pm}_i) \vert \over N_{\mu}}$,
where $N_{\mu}$ is the number of muons; Angle between the
momenta of same electric charge sign (SS) muons pairs; Average per event of the summed transverse 
momenta for leptons pairs  of same sign (SS) or opposite sign (OS),  $P_t^{SS,OS}(ll)=\sum_{(i,j)}
 {p_t(l_i^{\pm,\mp})+p_t(l_j^{\pm}) \over N}$, where $N$ is the number of configurations and
 $l=e,\mu $.

We have generated the inos single  production and the  $\tchi^0_1$
 pair production in separate samples of 5000 events each. Our choice of
 using equal number of events  for both reactions
has been made on the basis of the following three 
somewhat qualitative considerations, none of which is compelling.
 First, the single production reactions occur in company of their charge conjugate partners, which multiplies rates by a factor 2. Second, the other $\tilde \nu$ 
and $\tilde l$ single production reactions, which have not 
been included, would be expected to add contributions of
similar size to the leptonic distributions. Third, assuming 
for the RPV coupling constant the alternative
bound, $\l _{mJJ} { 100 GeV \over m_{\tilde l_R}  } <0.05$, there would  result a 
relative enhancement for single production  over pair 
production by a factor, $ ({ m_{\tilde l_R} \over 100 GeV }) ^2$.
The above three points motivate our rough guess that 
the number of events  chosen for the five single production processes 
(together with their charge conjugate partners) 
should be of comparable size to that of the $\tchi^0_1$ pair production process.

The results are shown in Figure \ref{MC}. 
The single production reactions present 
certain clear  characteristic features: Concentration of 
missing energy $E_m$ at low energies; pronounced peaks in the muon transverse 
momentum $P_t(\mu^{\pm})$ 
and in the angular distribution for the same sign muons pairs;  and 
a double peak in the transverse 
momentum distribution for the opposite sign leptons pairs,
$P^{OS}_t(ll)$. The large transverse 
momentum components  present in the single production distributions in
$P^{OS}_t(ll)$ and in $P_t(\mu^{\pm})$
 are explained by the fact that
 one of the two  leptons (namely, $l^{\pm}_m$)  is created at the production stage. Similarly, the existence of a strong
 angular correlation between same sign muons pairs is interpreted naturally
 by the momentum conservation balance between the 
 lepton $l^{\pm}_m$ produced in the initial stage and 
the other lepton produced at the decay stage. 
Although there are certain distinguishing properties
 between the single and pair production processes, the discrimination between the two may depend
 crucially on the relative sizes of the associated event samples. 
 Of course, the best  possible situation would
 be for an energetically forbidden neutralino pair production.

Finally, we comment on the effect of eventually excluding the $\tchi^0_2$ single
production component. In that case, most of the signals for single production would 
become less diluted in comparison with neutralino pair production, the large missing 
energy signal would be removed while the large OS lepton pair $P^{OS}_t(ll)$ signal
would become amplified.

\section{Conclusions}

\label{sectionC}

We have analysed the full set of $2 \to 2$ single production processes at leptonic colliders 
induced by the RPV interactions $LLE^c$, within a supergravity model. Although our
 approximate study has obvious limitations (factorisation and narrow resonance approximation,
 neglect of the spin correlations and omission of acceptance cuts), it uncovers the general
 trends of all the 5 single production reactions. Over the whole 
parameter space, for an RPV coupling constant $\l_{mJJ}$ of order $0.05$, 
the integrated rates are of comparable order
 of magnitudes although $\tchi_1^{\pm}$
 and $\tilde \nu$ are typically larger by factors of 2
 compared to $\tilde l$ production and   by 
factors of 5 compared to $\tchi_1^0$ production.
 The detectability for each single production
 separately is modest at LEPII  but comfortable at 
NLC, corresponding to a few events and a few thousands of events
 per year, respectively. 
A wide region of the parameter space
can be probed  at  $\l_{mJJ} > 0.05$. 
 In spite of the rich variety of final states, 
the dominant signals arise from the single
 or double RPC induced cascade decays to the LSP, which are 
also favored by phase space arguments. 
For the minimal supergravity model, assuming electroweak symmetry breaking,
 large mass differences occur between the scalar
 superpartners and the $\tchi_1^0$ LSP, leading to dominant 
cascade decays modes with weak competitivity from
 the RPV direct decays. The signals, 4l+$\Eslash $,6l,6l+$\Eslash $,
 arising from cascade decays are 
free from standard model background which make them quite interesting signatures 
for the discovery of supersymmetry and 
R parity violation. Even the 4l signal arising from direct RPV decays
 could be observable due to a
characteristic non diagonal flavor configuration. For center of mass 
energies well above all the
 thresholds, the 4l+$\Eslash $ signal receives contributions from
 all five single production processes and hence should
 be strongly amplified. We have presented some dynamical distributions for 
the final states which could characterise the single production reactions.

\section{Acknowledgments}

We are grateful to R. Barbier and S. Katsanevas for 
guidance in using the events generator SUSYGEN,
to P. Lutz and the Direction of the DAPNIA at Saclay  
for providing  us, through the 
DELPHI Collaboration,  an  access to  the ANASTASIE 
computer network at Lyon and  to P. Micout for his technical 
help on  that matter. 
We thank M. Besan\c con, F. Ledroit, R. Lopez and 
G. Sajot for helpful discussions. 
\clearpage

\appendix

\renewcommand{\thesubsection}{A.\arabic{subsection}}
\renewcommand{\theequation}{A.\arabic{equation}}
\setcounter{subsection}{0}
\setcounter{equation}{0}

\section{Formulas for spin summed amplitudes}

\label{seca}

We  discuss the five $2 \to 2$ body single production processes 
given by eq.(\ref{eqrc1}).  
The  formulas for the probability amplitudes  are:
\begin{eqnarray}
M(\tchi_a^-+ l^+_m)&=& {g\l_{mJJ} V^\star _{a1} \over s-m^2_{\tilde \nu_{mL}} } 
\bar v(k') P_L u(k) \bar u^c (p)P_Lv(p') 
- {g\l_{mJJ}V^\star _{a1}\over t-m^2_{\tilde \nu_{JL} } } 
\bar u^c (p) P_L u(k) \bar v(k') P_L v(p') ,\cr 
M(\tchi_a^0+ \bar \nu_m)&=& + { \sqrt 2 g\l_{mJJ} \over 
s-m^2_{\tilde \nu _{m L} }  } 
{1 \over 2}(N^*_{a2}-tg\t_W N^*_{a1})\bar u(p) P_L v(p') \bar v(k') P_L u(k)\cr
&&+{ \sqrt 2 g\l_{mJJ} \over t-m^2_{\tilde l_{JL} }   } 
{1 \over 2}(N^*_{a2}+tg\t_W N^*_{a1})\bar u(p) P_L u(k) \bar v(k') P_L v(p')\cr
&&+{ \sqrt 2 g\l_{mJJ} \over u-m^2_{\tilde l_{JR} }   } 
(tg\t_W N^\star _{a2})\bar v^c(p') P_L u(k) \bar v(k') P_L v(p),\cr
M(\tilde l_{mL}^-(p) +W^+(p'))&=&  { g \l^*_{mJJ} \over 
\sqrt 2 (s-m^2_{\tilde \nu_{m L} } )  } 2 p\cdot \e (p') 
 \bar v(k') P_R u(k) 
+ {g\l_{mJJ}^\star  \over \sqrt 2 t} \bar v(k') 
\g \cdot \e (p') (\pslash-\kslash ) P_R u(k) , \cr
M(\tilde \nu_{mL}(p) + Z(p'))&=&   {g \l^*_{mJJ} \over 2 \cos \t_W }
[{\bar v(k') \g \cdot \e (p') (\kslash-\pslash )  a_L(e) P_R u(k)
\over t-m^2_{ l_J } } \cr
&&+ {\bar v(k')a_R(e) P_R (\kslash-\pslash ')
 \g \cdot \e (p') u(k)\over u-m^2_{ l_J } } 
+ {\bar v(k')a_L(\tilde \nu) P_R u(k) 2p \cdot \e (p')
\over s-m^2_{ \tilde \nu_{mL} } } ], \cr
M(\tilde \nu_{mL}(p) + \g(p'))&=&- e \l^*_{mJJ} [{\bar v(k')\g \cdot \e (p')
(\kslash-\pslash ) P_R u(k) \over t-m^2_{ l_J } }  
+{\bar v(k') (\kslash - \pslash ' ) \g \cdot \e (p') P_R u(k) 
\over u-m^2_{ l_J } }]. 
\end{eqnarray}
In deriving the results for the inos  production  amplitudes,
we have systematically neglected their higgsino components.
The parameters in the $Z^0 f {\bar f}$ and $Z^0 \tilde f \tilde {\bar f}$ vertices  
denoted as, $a_H(f)=a(f_H)$ and $a_H(\tilde f)=a(\tilde f_H)$, 
are defined by, $a(f_H)=a(\tilde f_H)=2T_3^H(f)-2Qx_W$, 
with $H=[L,R]$ and $x_W=\sin ^2 \t_W$. Throughout this work, 
our notations follow closely the Haber-Kane conventions \cite{haber}.

The  unpolarized cross sections in the 
center of mass frame are given by  the familiar formula, 
$d\s /d\cos \t =p /(128 \pi k s) \sum_{pol} \vert M\vert^2 $,  where the sums over 
polarizations  for the probability amplitudes squared are given by:

\begin{eqnarray}
\sum_{pol} &&\vert M(\tchi^-_a +l_m^+ )  \vert^2 = 
 \vert \l_{mJJ} g V_{a1}^\star \vert^2  
\bigg [ { s(s-m_{\tchi^-_a}^2-m_{l_m}^2)  \over \vert R_s(\tilde \nu_{mL} )  \vert^2}
+ { (m_{\tchi^-_a}^2-t)(m_{l_m}^2-t) 
\over \vert R_t(\tilde \nu_{JL} ) \vert^2} \cr &-& 
Re \bigg ( { (s(s-m_{\tchi^-_a}^2-m_{l_m}^2) +
(m_{\tchi^-_a}^2-t)(m_{l_m}^2-t) - (m_{\tchi^-_a}^2-u)(m_{l_m}^2-u))  
\over R_s(\tilde \nu_{mL} ) R^\star _t 
(\tilde \nu_{JL}) } \bigg ) \bigg ],
\label{eqrc4a}
\end{eqnarray}

\begin{eqnarray}
\sum_{pol} &&\vert M(\tchi^0_a +\bar \nu_m )  
\vert^2 = { g^2 \over 2 } \vert \l_{mJJ}\vert ^2 
 \bigg [ \vert N_{a2}+tg\t_W N_{a1} \vert ^2 {t(t-m_{\tchi^0_a}^2) \over 
\vert R_t(\tilde l_{JL} )\vert^2 }     \cr
&+&4 \vert tg\t_W N_{a2} \vert ^2 { u(u-m_{\tchi^0_a}^2) \over 
\vert R_u(\tilde l_{JR} )\vert^2 }
+ \vert N_{a2}-tg\t_W N_{a1} \vert ^2 { s(s-m_{\tchi^0_a}^2) \over 
\vert R_s(\tilde \nu_{mL})\vert^2  }  \cr
&-& Re \bigg ( (N^*_{a2}-tg\t_W N^*_{a1})(-N_{a2}-tg\t_W N_{a1}) {
(s(s-m_{\tchi^0_a}^2) -t
(m_{\tchi^0_a}^2-t) +u (m_{\tchi^0_a}^2-u))
 \over 
 R_s(\tilde \nu_{mL}) R^*_t(\tilde l_{JL} )} \cr
&+& 2  (N^*_{a2}-tg\t_W N^*_{a1})(-tg\t_W N_{a2}) {
(s(s-m_{\tchi^0_a}^2)-u(m_{\tchi^0_a}^2-u)+t(m_{\tchi^0_a}^2-t))
 \over 
R_s(\tilde \nu_{mL})R^*_u(\tilde l_{JR}) } \cr
&+&2  (-N^*_{a2}-tg\t_W N^*_{a1})(-tg\t_W N_{a2}) {
(-u(m_{\tchi^0_a}^2-u)-t(m_{\tchi^0_a}^2-t)-s(s-m_{\tchi^0_a}^2))
 \over 
R_t(\tilde l_{JL} )R^*_u(\tilde l_{JR}) } \bigg ) \bigg ],
\label{eqrc4b}
\end{eqnarray}

\begin{eqnarray}
\sum_{pol} &&\vert M(\tilde l^-_{mL} +W^+)\vert ^2 =
{s g^2 \vert \l_{mJJ}\vert ^2  \over 2 \vert R_s(\tilde \nu_{mL}) \vert ^2 } 
  ( { (s-m^2_{\tilde {l}^-_{mL}}-m_W^2)^2\over m_W^2} -4 m^2_{\tilde {l}^-_{mL}} ) 
- {g^2 \vert \l_{mJJ}\vert ^2  \over  2 \vert t \vert ^2 }\cr & \times & 
[(m^2_{\tilde {l}^-_{mL}}-t)(m_W^2-t) +st  
+{m_W^2-t\over m_W^2} 
\bigg ( (m^2_{\tilde {l}^-_{mL}}-t)(m_W^2+t) + t(m_W^2-u)\bigg )  ] \cr
&-& g^2 Re {\l_{mJJ} \l ^\star _{mJJ}  \over t R^*_s(\tilde \nu_{mL})   } 
[(m^2_{\tilde {l}^-_{mL}}-t)(m^2_{\tilde {l}^-_{mL}}-u) +s (m^2_{\tilde {l}^-_{mL}}-u)+ 
 (m^2_{\tilde {l}^-_{mL}}-t)(m^2_W-t)\cr
&-& {s(s-m^2_W-m^2_{\tilde {l}^-_{mL}})(m^2_W-t)\over m_W^2} ],
\label{eqrc4c}
\end{eqnarray}

\begin{eqnarray}
\sum_{pol} &&\vert M(\tilde \nu_{mL} + Z)\vert ^2 ={g^2 \vert \l_{mJJ}\vert ^2
 \over  \cos^2\t_W } Re \bigg [{s \over \vert R_s(\tilde \nu_{mL})\vert^2 }   
( { (s-m_{\tilde \nu_{mL}}^2-m_Z^2)^2\over 4 m_Z^2} -m^2_{\tilde \nu_{mL}} ) \cr
&-&{( \sin^2\t_W)^2 \over  \vert R_u(l_J)\vert^2}
\bigg ( (m_{\tilde \nu_{mL}}^2-u)(m_Z^2-u) + su 
+{ m_Z^2-u \over m_Z^2} 
 ( (m_{\tilde \nu_{mL}}^2-u)(m_Z^2+t) + u (m_Z^2-t)  ) \bigg ) \cr & - &
{(2 \sin^2\t_W-1)^2 \over 4 \vert R_t(l_J)\vert^2 } \bigg (  (m_{\tilde \nu_{mL}}^2-t)(m_Z^2-t) + st+{m_Z^2-t \over m_Z^2} 
 ( (m_{\tilde \nu_{mL}}^2-t)(m_Z^2+t) + t (m_Z^2-u) )    \bigg ) \cr  
&-&  { ( \sin^2 \t_W) \over   R_u(l_J)  R^*_s(\tilde \nu_{mL}) } 
\bigg ( (m_{\tilde \nu_{mL}}^2-t)(m_{\tilde \nu_{mL}}^2-u)+s (m_{\tilde \nu_{mL}}^2-t)   
+ (m_{\tilde \nu_{mL}}^2-u)(m_Z^2-u) \cr
&-&{s\over m_Z^2}(s-m_{\tilde \nu_{mL}}^2-m_Z^2)(m_Z^2-u) 
\bigg ) 
+ {(2 \sin^2\t_W-1)\over  2 R_t(l_J) R^*_s(\tilde \nu_{mL})}
 \bigg ( (m_{\tilde \nu_{mL}}^2-t)(m_{\tilde \nu_{mL}}^2-u) +s(m_{\tilde \nu_{mL}}^2-u) \cr
&+& 
(m_Z^2-t)(m_{\tilde \nu_{mL}}^2-t)  -{s\over m_Z^2} 
(m_Z^2-t)(s-m_{\tilde \nu_{mL}}^2-m_Z^2) \bigg )  
+ {(2 \sin^2\t_W-1)( \sin^2\t_W)\over R_t(l_J) R^*_u(l_J)} \cr && \times
\bigg ( (m_{\tilde \nu_{mL}}^2-u)(m_{\tilde \nu_{mL}}^2-t) + s \ m_{\tilde \nu_{mL}}^2 
-{1 \over M_Z^2}(-{(s-m_{\tilde \nu_{mL}}^2-m_Z^2)\over 2}((m_{\tilde \nu_{mL}}^2-u)(m_Z^2-u)
 \cr &+& (m_{\tilde \nu_{mL}}^2-t)(m_Z^2-t)
- s (s-m_{\tilde \nu_{mL}}^2-m_Z^2))+(m_Z^2-u)(m_Z^2-t)m_{\tilde \nu_{mL}}^2)
\bigg )  \bigg ],
\label{eqrc4d}
\end{eqnarray}

\begin{eqnarray}
\sum_{pol} &&\vert M(\tilde \nu_{mL} + \g)\vert ^2 =
 2 e^2\vert \l_{mJJ}\vert ^2((m_{\tilde \nu_{mL}}^2-t)(m_{\tilde \nu_{mL}}^2-u)-sm_{\nu_m}^2)
[{1 \over \vert R_t(l_J) \vert ^2}
+{1 \over \vert R_u(l_J) \vert ^2}] \cr
&+& 4 e^2 \vert \l_{mJJ} \vert ^2 Re { (m_{\tilde \nu_{mL}}^2-t)
(m_{\tilde \nu_{mL}}^2-u) \over R_t(l_J) R_u^*(l_J)},
\label{eqrc4e}
\end{eqnarray}

where $ Re $ stands for the real part,  $ R_s(\tilde \nu_i) =s-m^2_{\tilde \nu_{i} }
+i m_{\tilde \nu} \Gamma_{\tilde \nu}, \ R_t(\tilde \nu_i ) =t-m^2_{\tilde \nu_{i} }$  
and  $R_u(\tilde \nu_i ) =u-m^2_{\tilde \nu_{i} },   
\ [s= (k+k')^2, \ t=(k-p)^2, \ u= (k-p')^2] $, with similar 
definitions applying for the propagator factors $R_{s,t,u}(l_i,\tilde l_i)$.

\section{Formulas for partial decay widths}

\label{secappa}

\renewcommand{\thesubsection}{B.\arabic{subsection}}
\renewcommand{\theequation}{B.\arabic{equation}}
\setcounter{subsection}{0}
\setcounter{equation}{0}

The formulas for the various two-body decay widths are quoted below. 

\begin{eqnarray}
\G (\tilde \nu \to \tchi^+_a +l^-)= {g^2\over 16 \pi }
\vert V_{a1}\vert ^2 m_{\tilde \nu } 
\bigg (1-{m^2_{\tchi_a} \over m^2_{\tilde \nu } }\bigg )^2
\label{eqpr4}
\end{eqnarray}
\begin{eqnarray}
\G (\tilde \nu \to \tchi_a^0 +\nu )= {g^2\over 32 \pi }
\vert N_{a2} -N_{a1} \tan \t_W \vert ^2 m_{\tilde \nu } 
\bigg (1-{m^2_{\tchi_a} \over m^2_{\tilde \nu } }\bigg )^2
\label{eqpr5}
\end{eqnarray}
\begin{eqnarray}
\G (\tilde l^+_L \to \tchi^+_a \bar \nu)= {g^2\over 16 \pi }
\vert U_{a1}\vert ^2 m_{\tilde l_L } 
\bigg (1-{m^2_{\tchi_a} \over m^2_{\tilde l_L } }\bigg )^2
\label{eqpr6}
\end{eqnarray}
\begin{eqnarray}
\G (\tilde l^-_{[L,R]} \to \tchi_a^0 +l^- )= {g^2\over 32 \pi }
[\vert N_{a2} +N_{a1} \tan \t_W \vert ^2, \vert N_{a2} \tan \t_W \vert^2] 
m_{\tilde l_H} 
\bigg (1-{m^2_{\tchi_a} \over m^2_{\tilde l_H } }\bigg )^2
\label{eqpr7}
\end{eqnarray}

\begin{eqnarray}
\G (\tilde \nu_i (M) \to l^-_k (m_1) +l^+_j (m_2)) 
&=&\G (\tilde l^-_{jL} (M) \to \bar \nu_i (m_1) +l^-_k (m_2)) \cr 
&=&\G (\tilde l^-_{kR} (M) \to \nu_i (m_1) +l^-_j (m_2)) \cr 
&=& {\vert \l _{ijk}\vert^2 \over 8 \pi } 
k (1-{m_1^2+m_2^2\over M^2 } ) 
\label{eqpr8}
\end{eqnarray}
\begin{eqnarray}
\G (\tchi^{\pm}_m (M_{\pm}) \to \tchi_l^0  (M_0) +W^{\pm} (m_W) )&=& {g^2 \vert k \vert \over  
16\pi  M_{\pm}^2} \bigg [(\vert O_L\vert^2 + \vert O_R\vert^2 ) 
\bigg ( (M_+^2+M_0^2-m_W^2) \cr && +{1\over m_W^2} (M_{\pm}^2-M_0^2-m_W^2) 
(M_{\pm}^2-M_0^2+m_W^2) \bigg ) \cr && -
12  M_0 M_{\pm} Re\bigg (O_L O_R^\star \bigg )  \bigg ]
\label{eqpr9}
\end{eqnarray}
\begin{eqnarray}
\G (\tchi^0_a\to \tilde f_{[L,R]} \bar f' )&=& {g^2 M_0\over  16\pi }
(1-{m^2_{\tilde f} \over M^2_0})^2 \cr &&
[\vert T^f_3 N_{a2} - \tan \t_W (T^f_3 - Q^f) N_{a1} \vert ^2 , 
\vert \tan \t_W  N_{a2} \vert ^2 ]
\label{eqpr10}
\end{eqnarray}
\begin{eqnarray}
\G (\tchi^{\pm}_a\to \tilde f_{[T^f_3=-1/2, 1/2]} \bar f' )&=& 
{g^2 M_{\pm}\over 32 \pi }
(1-{m^2_{\tilde f} \over M^2_{\pm}})^2 [\vert U_{a1} \vert ^2, 
\vert V_{a1} \vert ^2].
\label{eqpr11}
\end{eqnarray}

We use the notations: $ O^L=O^L_{lm}= N_{l2} V_{m1}^\star -{1\over \sqrt 2} 
N_{l4} V_{m2}^\star, \ O^R=O^R_{lm}= N_{l2} U_{m1} +
{1\over \sqrt 2} N_{l3}^\star  U_{m2}, \ M_{\pm}=m_{\tchi^{\pm}_a}, \
M_0=m_{\tchi^0_a}$ and $ k= \l^\ud (M^2, m_1^2,m_2^2)/2M$ 
with $\l(a,b,c)=a^2+b^2+c^2+ab+bc+ac$. The notations, $T_3^f, \ Q^f$, stand for the third 
component of the $SU(2)_L$ group and the electric charge of the fermion $f$.
We have omitted the higgsino components of the inos.
We shall use the simplified formulas for the RPC three-body decays, 
$ \tchi_m^-\to \tchi_l^0 +l\bar \nu , \ q \bar q  $,  obtained by neglecting the 
three-momenta in the W and $\tilde l$ propagators, as  quoted in \cite{fe}. We have set
 in these formulas, the flavor and color parameters to, $N_f=2, \ N_c=3$ for quarks and
 $N_f=3, \ N_c=1$ for leptons.
The formulas for the spin summed amplitudes of the 
RPV decays   $\tchi^-_a\to \bar \nu_i \bar \nu_j l_k^- , \
\tchi^-_a \to  l_k^+ l_j^- l_i^- $, associated to the coupling constants $\l_{ijk}$, were first 
derived in  \cite{fourjet1} (see the appendix). The integrated decay rates are given by 
familiar formulas \cite{PartData} involving twofold integrals over the
final state three-body phase space. 
If we neglect the final particles masses, an analytic formula can be 
derived for the integral giving the contributions to the charginos partial 
rates associated with the gauginos components only 
(neglecting the higgsino components contribution). For completeness, we display 
the final results:

\begin{eqnarray}
\G (\tchi^-_a)&=&M_{\tchi^-_a} {g^2 X^2_{a1} \vert \l _{ijk}\vert^2 \over 128 \pi^3} \bigg
[{1 \over8 } \bigg (-5+6\mu_i+(2-8 \mu_i+6 \mu^2_i) \log (1-{1 \over \mu_i}) \cr 
&-&5+6\mu_j+(2-8 \mu_j+6 \mu^2_j) \log (1-{1 \over \mu_j}) \bigg ) \cr
&+& {1 \over 2 } \bigg (  \mu_i + \mu_j -{1 \over 2} 
+(\mu^2_i- \mu_i) \log (1-{1 \over \mu_i})+(\mu^2_j- \mu_j) \log (1-{1 \over \mu_j}) \cr
&-& \mu_i \mu_j \log (1-{1 \over \mu_i}) \log ({ \mu_i +\mu_j-1\over \mu_j }) \cr
&-& \mu_i \mu_j \log (1-{1 \over \mu_j}) \log ({ \mu_i +\mu_j-1\over \mu_i }) \cr
&+& \mu_i \mu_j  [ Sp({ \mu_i\over \mu_j})+Sp({ \mu_j\over \mu_i})
- Sp({ 1-\mu_i\over \mu_j}) - Sp({ 1-\mu_j\over \mu_i}) ]\bigg ) \bigg ],
\label{eqpr111}
\end{eqnarray}

where $Sp(x)=Polylog(x)=Li_2(x)$ is the Spence or Polylog function. We use the notations 
$\mu_{\a}= m^2_{\tilde \nu_{\a}}/M^2_{\tchi^-_a}, \ [\a =i,j] $, $X_{a1}=U_{a1}$ for the decay
$\tchi^-_a\to  l_k^+ l_j^- l_i^-$, and  $\mu_{\a}= m^2_{\tilde l_{\a}}/ M^2_{\tchi^-_a}, 
 \ [\a =i,j] $, $X_{a1}=V_{a1}$ for the decay $\tchi^-_a\to  \bar \nu_i \bar \nu_j l_k^- $.

\end{document}